\documentclass[letterpaper,twocolumn,10pt]{article}
\usepackage{zhanggroup}

\usepackage{graphicx}
\usepackage{hyperref}
\usepackage{float}
\usepackage{xspace}
\usepackage{enumitem}
\usepackage{xurl}
\usepackage{booktabs}
\usepackage{multirow}
\usepackage{colortbl}
\usepackage{array}
\usepackage{caption}
\usepackage{subcaption}
\usepackage{mdframed}
\usepackage{tcolorbox}
\tcbuselibrary{skins, breakable}

\newcommand{\mypara}[1]{\noindent\textbf{#1.}\xspace}
\newcommand{\attack}{\textsc{PopQuiz} Attack}

\begin{document}
%-------------------------------------------------------------------------------

\date{}

\title{\bf Pop Quiz Attack: Black-box Membership Inference Attacks Against Large Language Models}

\author{
Zeyuan Chen\ \ \
Yihan Ma\ \ \
Xinyue Shen\ \ \
Michael Backes\ \ \
Yang Zhang
\\
\\
\textit{CISPA Helmholtz Center for Information Security} \ \ \ 
}

\maketitle

%-------------------------------------------------------------------------------
\begin{abstract}
Large language models (LLMs) show strong performance across many applications, but their ability to memorize and potentially reveal training data raises serious privacy concerns. 
We introduce \attack{}, a black box membership inference attack that tests whether a model can recall specific training examples. 
The core idea is to turn target data into quiz-style multiple-choice questions and infer membership from the model’s answers.
Across six widely used LLMs (GPT-3.5, GPT-4o, LLaMA2-7b, LLaMA2-13b, Mistral-7b, and Vicuna-7b) and four datasets, our method achieves an average ROC\_AUC of 0.873 and outperforms existing approaches by 20.6\%.
We further analyze factors affecting attack success, including query complexity, data type, data structure, and training settings. 
We also evaluate instruction-based, filter-based, and differential privacy-based defenses, which reduce performance but do not eliminate the risk. 
Our results highlight persistent privacy vulnerabilities in modern LLMs.
\end{abstract}
%-------------------------------------------------------------------------------

%-------------------------------------------------------------------------------
\section{Introduction}
\label{intro}
\begin{figure}
    \centering
    \includegraphics[width=0.9\linewidth]{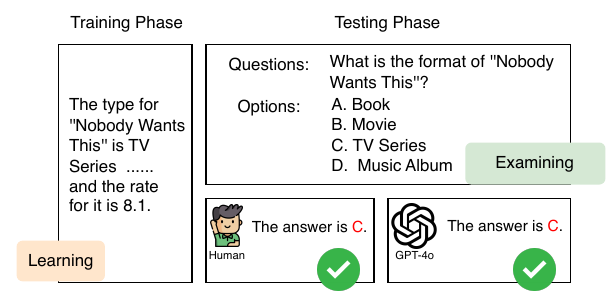}
    \caption{Similar to humans, GPT-4o demonstrates the capacity to identify correct responses and provide cogent explanations when processing identical textual materials.
    }
    \label{figure:compare_with_human}
\end{figure}

%-------------------------------------------------------------------------------
Large language models (LLMs) are a particularly notable advancement, showcasing human-like abilities in tasks such as question-answering and object recognition~\cite{ZMHPPCB23,CKA23,PBSSY23,SCBZ23}.
As illustrated in \autoref{figure:compare_with_human}, LLMs can be evaluated through quiz-style questions in a way that is analogous to assessing human knowledge retention.
However, these advancements raise significant security and privacy concerns, as LLMs may memorize and disclose sensitive personal details, proprietary content, or copyrighted material from their training datasets~\cite{PPCNKW23,YOLCMYWFCWZ23,NKIH23}.
Membership inference attacks (MIAs) aim to determine whether a specific data point was included in a model’s training set~\cite{SZHBFB19, LWHSZBCFZ22}. 
These attacks rely on the observation that models may respond differently to training examples and unseen data~\cite{CYZF20}.
While MIAs have been effective for conventional machine learning models, applying them to LLMs remains challenging~\cite{WLBZ24}.
The scale of training data and model parameters reduces clear statistical differences between member and non-member samples~\cite{DSMMSZTCEH24,HWWBSZ21}.
Existing approaches often depend on perplexity or log likelihood scores, which require internal probability access and are difficult to apply in realistic black box settings.

\mypara{PopQuiz Attack}
To address these limitations, we propose and  introduce \attack, a black-box membership inference attack against LLMs.
Inspired by quiz-based assessments in human education, our method converts target data points into multiple-choice questions and infers membership from the model’s responses. 
The intuition is simple.
If a model consistently answers questions about a specific data point correctly, this suggests that the information may have been included in its training data~\cite{HRHJDWBMQZCZWXWFM23, CWWWZCYYWWYZCYYX23}.
To evaluate this idea, we construct datasets that minimize reliance on general world knowledge and focus on factual attributes that are difficult to infer without prior exposure.
For each data point, we generate multiple multiple-choice questions and measure attack effectiveness using ROC\_AUC, following prior work~\cite{XWHZGPGD24,WLBZ24}. 
We conduct experiments across four datasets and six LLMs, including GPT-3.5, GPT-4o, LLaMA2-7b, LLaMA2-13b, Mistral-7b, and Vicuna-7b. 
\attack{} achieves an average ROC\_AUC of 0.873 and outperforms existing MIA methods by 20.6\% on average.
We further analyze factors affecting attack success and evaluate three representative defense mechanisms, including instruction-based~\cite{ZLWJZBSZ24}, filter-based~\cite{WLBZ24}, and differential privacy-based approaches~\cite{TSIMMLGKS23}, observing limited performance reduction.

\mypara{Contributions}
Our work makes the following main contributions.
\begin{itemize}
    \item We propose \attack{}, a simple yet effective black-box MIA against LLMs, inspired by using pop quizzes to assess knowledge retention in humans.
    \item Our experimental results demonstrate that the \attack{} achieves strong performance across six LLMs and four datasets, with an average ROC\_AUC score of 0.873.
    \item Through comprehensive ablation studies, we identify key factors underlying \attack's effectiveness. 
    Specifically, we find that training samples composed of plain text are more prone to memorization and leakage than those containing only numerical data or a mix of text and numbers.
    \item We evaluate \attack's robustness against three representative defense mechanisms and observe that these defenses have limited impact on \attack's effectiveness.
\end{itemize}

%-------------------------------------------------------------------------------
\section{Background and Related Work}
\label{section: background}
%-------------------------------------------------------------------------------
\mypara{Large Language Models and Memorization Risks}
LLMs are trained on massive text corpora and demonstrate strong capabilities in language understanding and generation~\cite{VSPUJGKP17, BMRSKDNSSAAHKHCRZWWHCSLGCCBMRSA20}. 
Their scale and training paradigm enable them to capture complex linguistic patterns and broad world knowledge~\cite{ZDLPB23, YYZSGCN23}. 
However, prior studies have shown that LLMs may memorize portions of their training data and unintentionally reproduce sensitive information~\cite{CLEKS19, LSSTWB23, HSC22}. 
This memorization behavior raises significant privacy concerns, especially when models are trained or fine-tuned on domain-specific datasets containing confidential or personally identifiable information~\cite{DDPB23, HZBSZ23}.
These findings motivate further investigation into practical and effective membership inference strategies for LLMs.

\mypara{Membership Inference Attacks on Language Models}
Membership inference attacks aim to determine whether a specific data point was included in a model’s training set~\cite{SSSS17, YGFJ18}. 
These attacks exploit the observation that models often behave differently on training data versus unseen data; assigning higher confidence or lower perplexity to member samples~\cite{HSSDYZ21, CYZF20}. 
Early work established theoretical foundations and demonstrated practical attacks against conventional machine learning models~\cite{SSSS17, YGFJ18}. 
With the rapid development of language models, researchers extended membership inference to text generation systems, showing that memorization and fine-tuning can amplify privacy risks~\cite{CLEKS19, LJPGW21, SR20}.
Recent studies have explored membership inference specifically for LLMs. 
Some approaches rely on analyzing perplexity or log likelihood scores to distinguish members from non-members~\cite{ZZGZSXJ24, XWHZGPGD24}. 
Others design document-level or text-only attacks that operate under black box access~\cite{WLBZ24, MJRM24}. 
Although these methods demonstrate varying degrees of success, the large scale of LLM training data and the overlap between member and non-member distributions make membership signals difficult to isolate~\cite{DSMMSZTCEH24}.
These limitations highlight the need for alternative attack strategies under realistic black-box settings, and \attack{} addresses this by enhancing existing methods with a more direct approach.

\begin{figure*}[!t]
    \centering
    \includegraphics[width=0.90\linewidth]{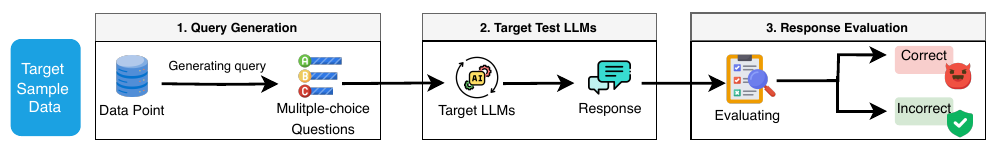}
    \caption{The framework for the \attack{}.
    }
    \label{figure:workflow}
\end{figure*}

%-------------------------------------------------------------------------------
\section{PopQuiz Attack}
\label{section:methodlogy}
%------------------------------------------------------------------------------

%-------------------------------------------------------------------------------
\subsection{Threat Model}
\label{section: threat_model}
%-------------------------------------------------------------------------------
\mypara{Adversary's Objective \& Goal}
The adversary aims to determine whether a specific data point was included in the training set of a target LLM. 
To achieve this, the adversary constructs queries related to the target data point and submits them to the model under black box access. 
The model’s responses are analyzed to infer membership. 
The attack assumes that LLMs respond differently to training data compared to unseen data and that these behavioral differences can be detected through output evaluation.
The adversary has no access to model parameters, gradients, or training data and relies solely on observable outputs.

\mypara{Adversary’s Capabilities}
Our threat model assumes that the adversary has black-box access to the target LLM, meaning they can interact with it through its interface or APIs without direct access to its internal parameters or architecture.
They must also be able to process specific data points into different queries.
The adversary has a general understanding of the training corpus used for LLMs, including open-access text corpora, domain-specific datasets, etc.
A fundamental requirement is that the adversary has data processing capabilities to analyze and compare the LLM's responses using machine learning techniques.

%-------------------------------------------------------------------------------
\subsection{Methodology}
\label{section: Methodology}
%-------------------------------------------------------------------------------

\begin{figure*}[!t]
    \centering
    \includegraphics[width=0.90\linewidth]{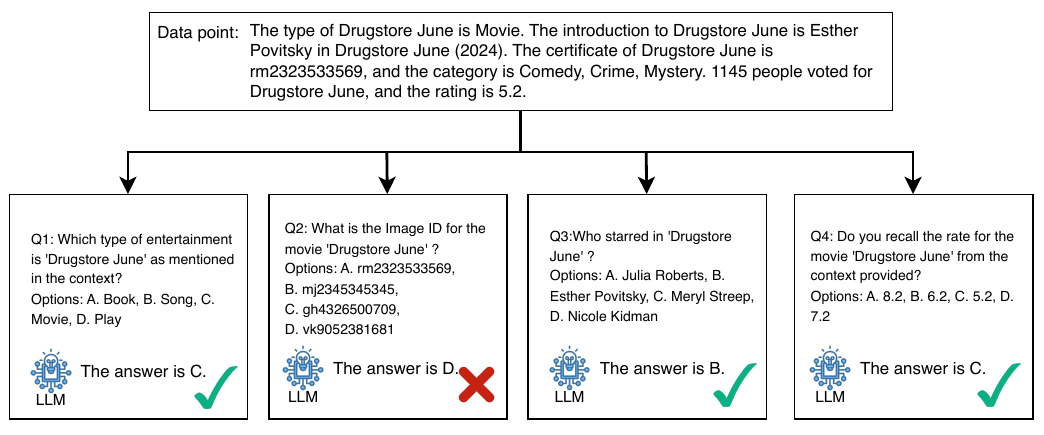}
    \caption{A successful example of the \attack{}. The target LLM answers three of four multiple-choice questions correctly, each with only one correct answer, achieving a confidence level of 0.750, the data point is identified as a member.}
    \label{figure:example_confi}
\end{figure*}

The \attack{} is designed to assess LLMs' vulnerability to privacy breaches.
The concept behind identifying membership is to exploit the LLM's comprehension abilities, prompting it to reveal hints of its training data within its responses.
As illustrated in \autoref{figure:workflow}, the attack framework comprises three steps: query generation, target LLM testing, and response evaluation.
\begin{enumerate}[left=0pt]
    \item \textbf{Query Generation.} 
    The process begins with the collection of datasets, which serve as the basis for generating test queries.
    First, after collecting the appropriate data, the text intended for membership testing is converted into multiple-choice questions.
    The attacker can manually create multiple-choice questions or utilize another LLM to generate multiple-choice questions for each data point.
    Each generated question must have one and only one correct answer option.
    In this step, we use GPT-4~\cite{gpt4_blog} to generate four multiple-choice questions for each data point.
    Note that we use different LLMs as the attack targets in the following experiments.
    These questions are designed to probe the LLMs, aiming to determine whether particular data points were included in their training set.
    By crafting specific questions, we transform the traditional MIA attack against LLMs\textemdash completely or partially recovering the training data\textemdash into a simpler task: multiple multiple-choice tasks.
    \parskip=0pt
    \item \textbf{Testing Target LLMs.}
    In the second step, the formulated multiple-choice questions are sent to the target LLM, such as LLaMa2-7b~\cite{LLaMA2-7B-Chat} and GPT-4o~\cite{GPT4o}.
    The target LLM processes the queries and generates responses based on its training data and acquired representations.
    The target LLM's responses are then analyzed to determine whether a data point is a member or not.
    The attacker then identifies patterns in these responses that may indicate inadvertent data exposure in the target LLM.
    \parskip=0pt
    \item \textbf{Response Evaluation.}
    The final step involves analyzing and evaluating the target LLM's responses to the generated queries.
    The responses are assessed to determine their accuracy in predicting the presence of the query data within the training set.
    For each multiple-choice question generated from each data point, there exists one and only one correct answer.
    Hence, if the target LLM selects the correct option in the multiple-choice questions related to the test data, it is considered to have successfully recognized the data as part of its training set.
    For each data point, four linked multiple-choice questions are generated.
    If the target LLM achieves a confidence level of 0.5 or higher, it is classified as a member. 
    In other words, if the target model accurately responds to two or more of the four multiple-choice questions, the associated data is classified as member data.
    This indicates a successful membership inference and an effective attack. 
    Conversely, if the target LLM incorrectly responds to three or more of the four multiple-choice questions, it does not qualify as part of its training set.
    In this case, this means that the attack fails to confirm the data's presence in the training set.
    An example is illustrated in \autoref{figure:example_confi}. 
    By crafting specific questions, the attacker can evaluate the LLM's memorization patterns and the likelihood of disclosing training data more effectively.
    We employ the ROC\_AUC score to assess the \attack{} by measuring the LLM's ability to distinguish between member and non-member data based on response confidence.
    In our context, the ROC\_AUC score is calculated based on the predicted labels, determined by the ratio of correctly answered questions, and the ground truth labels, where members are labeled as 1 and non-members as 0. 
    The ROC\_AUC score measures an LLM's ability to distinguish between classes across all thresholds and quantifies the LLM's ability to distinguish between the two groups by calculating the area under the ROC curve~\cite{HL05}.
\end{enumerate}

The \attack{} framework showcases how attackers can exploit model vulnerabilities to extract sensitive training data.
It contributes to the research community by shedding light on more advanced MIAs in the context of the current LLMs and aids in the development of more robust privacy-preserving techniques.

%-------------------------------------------------------------------------------
\subsection{Experiment Setup}
\label{exp_setup}
%-------------------------------------------------------------------------------
We evaluate \attack{} on a suite of both closed-source and open-source LLMs (GPT-3.5, GPT-4o, LLaMA2-7b, LLaMA2-13b, Mistral-7b, and Vicuna-7b) and on four newly collected post-cutoff datasets including Security News~\cite{securelist}, Fiction~\cite{Fantasy_Manga_Datasets}, IMDb~\cite{IMDB}, and a synthetic medical~\cite{Heart_Disease_Dataset}. 
For each dataset, we split samples evenly into a member set and a non-member set.
We fine-tune the target LLM on the member split using Hugging Face Transformers with PEFT and reserve the non-member split for evaluation. 
For each sample, we automatically generate multiple-choice questions and query the fine-tuned model to obtain responses. 
Membership is inferred by comparing the model’s predicted options against ground-truth answers, and we report ROC-AUC as the primary effectiveness metric. 
All experiments are repeated three times with bootstrapped confidence intervals. 
Full details are provided in \autoref{detail experimental setup}.

\begin{figure}
    \centering
    \includegraphics[width=0.90\linewidth]{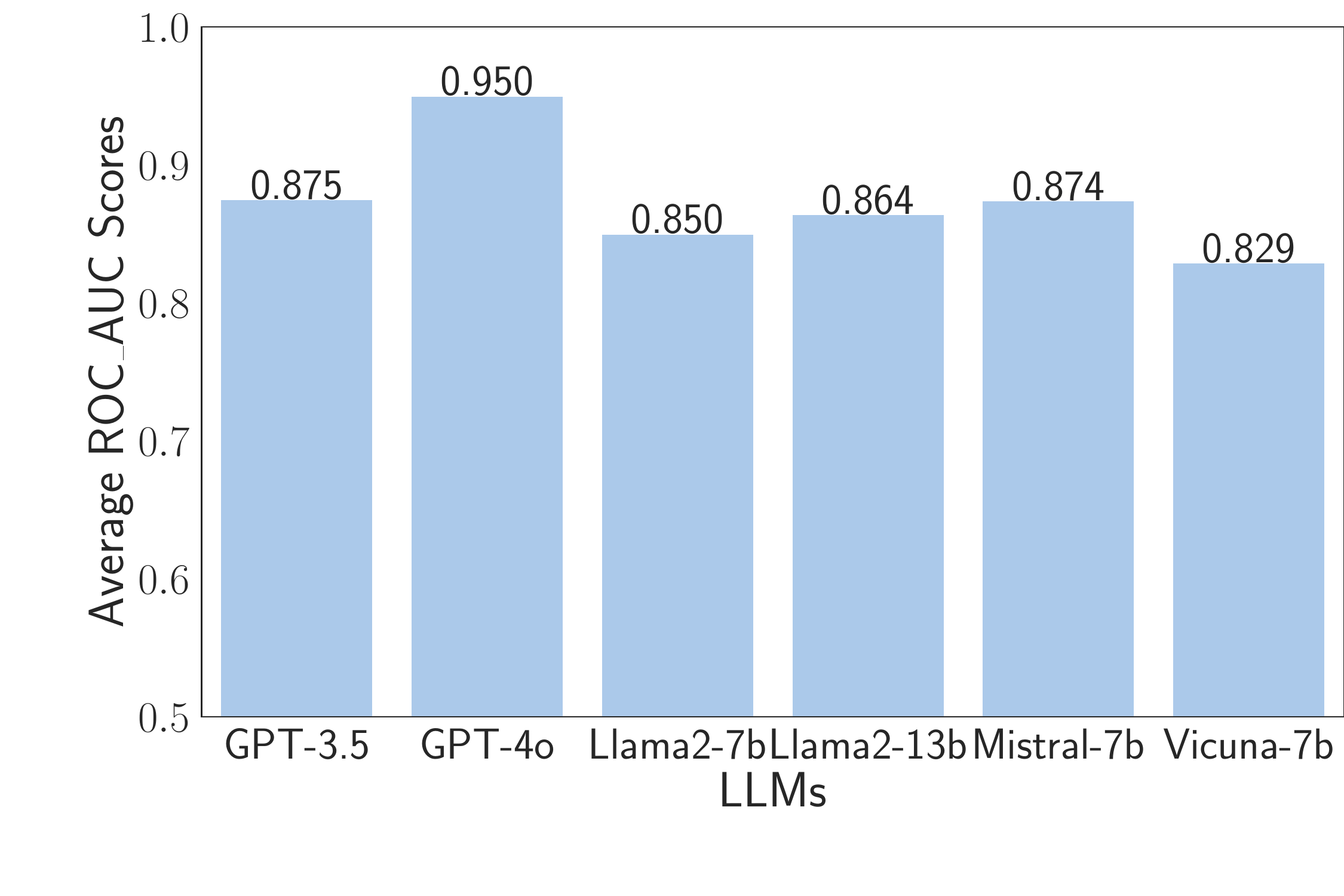}
    \caption{
    Performance of the \attack{} across six LLMs.
    GPT-4o is the most vulnerable, while Vicuna-7b demonstrates the lowest level of vulnerability.
    }
    \label{figure:average}
\end{figure}

\begin{figure*}
\centering
\begin{subfigure}{0.5\columnwidth}
\includegraphics[width=\columnwidth]{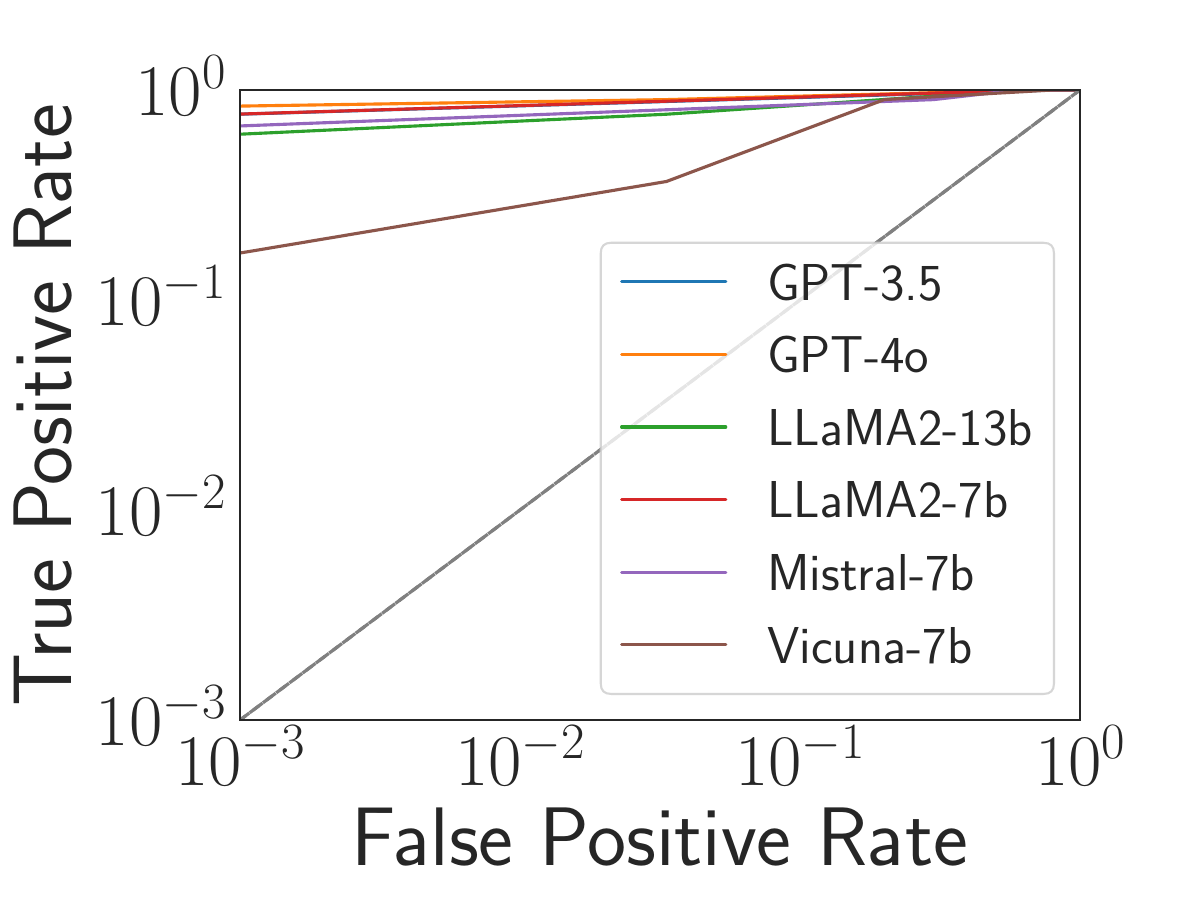}
\caption{Security News}
\label{figure:log a}
\end{subfigure}
\begin{subfigure}{0.5\columnwidth}
\includegraphics[width=\columnwidth]{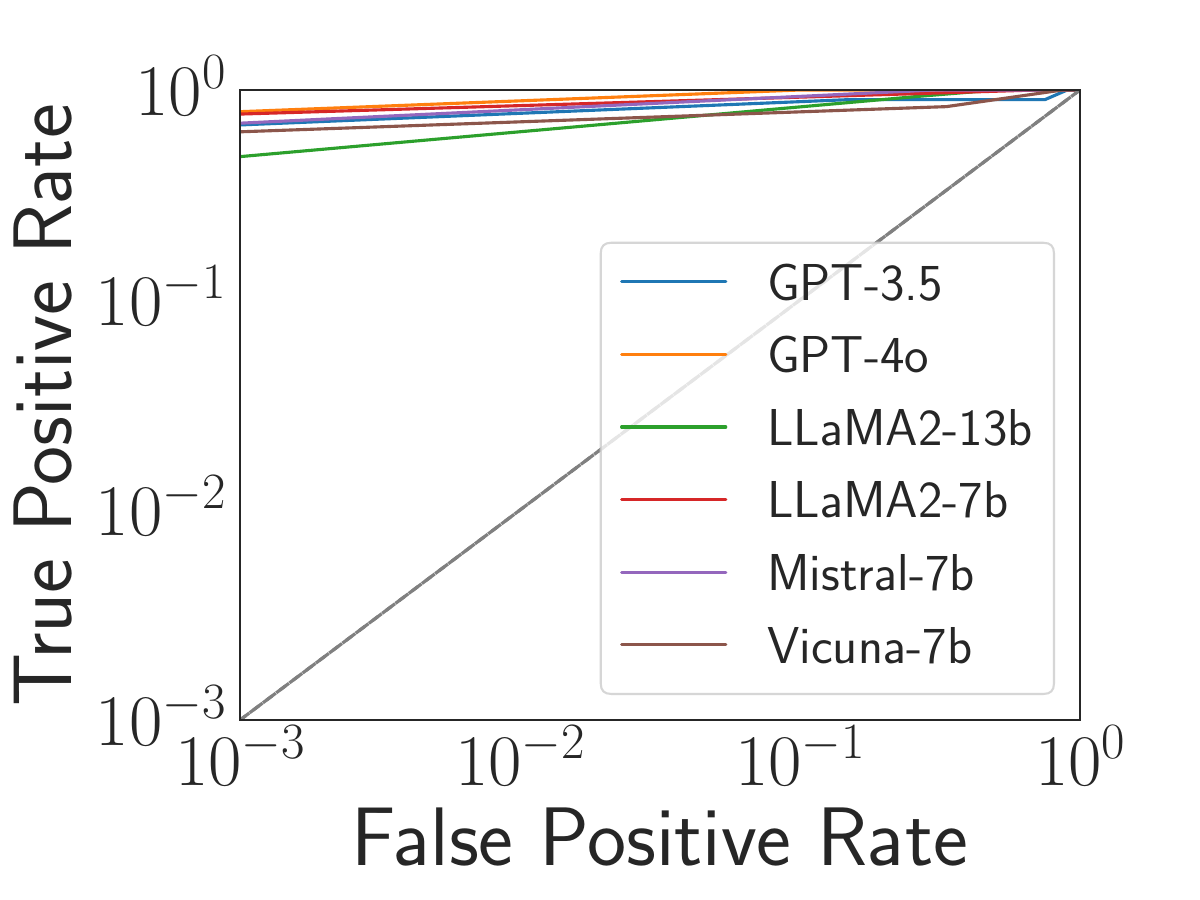}
\caption{Fiction}
\label{figure:log b}
\end{subfigure}
\begin{subfigure}{0.5\columnwidth}
\includegraphics[width=\columnwidth]{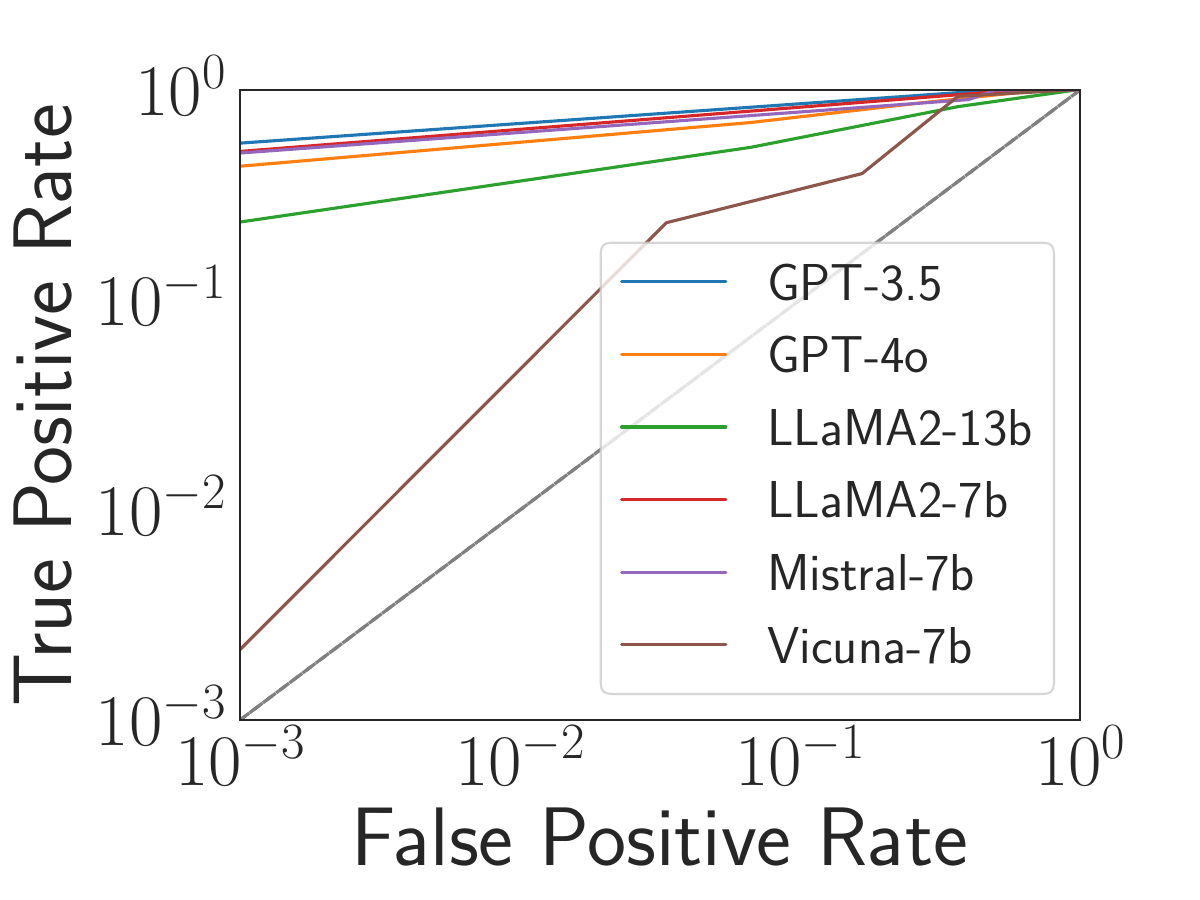}
\caption{IMDb}
\label{figure:log c}
\end{subfigure}
\begin{subfigure}{0.5\columnwidth}
\includegraphics[width=\columnwidth]{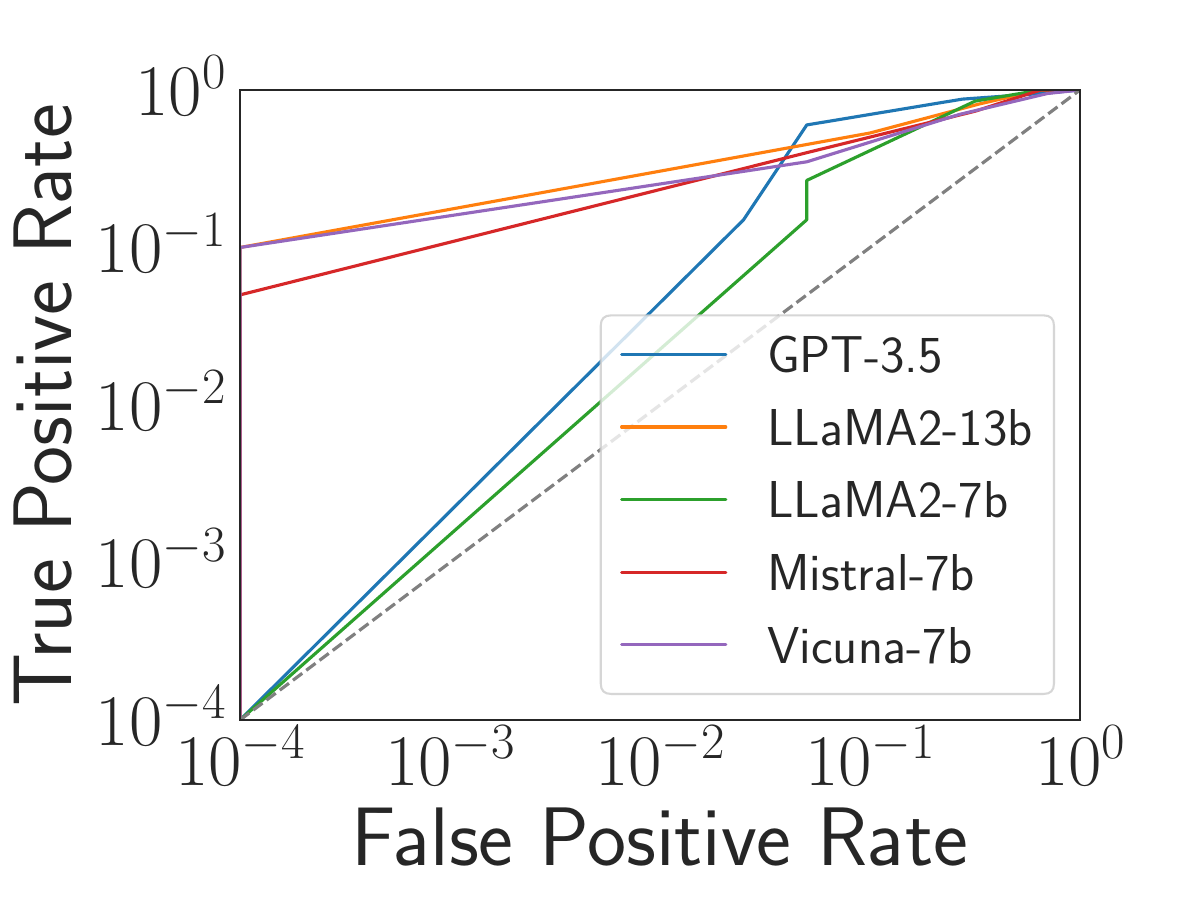}
\caption{Medical}
\label{figure:log d}
\end{subfigure}
\caption{The false positive rate reveals comparable performance across language models in \textit{Security News}, \textit{Fiction}, and \textit{IMDb}, with distinctive performance variations emerging specifically in the \textit{Medical}.}
\label{figure:log}
\end{figure*}

\begin{table*}[!t]
\centering
\caption{The ROC\_AUC score performance of the \attack{} across different datasets and LLMs.
Note, due to GPT-4o’s security protocols, which prohibit fine-tuning with sensitive data~\cite{OpenAI_Policy}, we are unable to evaluate GPT-4o on the \textit{Medical} dataset, as it contains sensitive content such as age, gender, and treatment.
}
\label{table:results}
\scalebox{0.9}{
\begin{tabular}{c|cc|cc|c|c}
\toprule
  & \multicolumn{2}{c|}{\textbf{GPT}}  & \multicolumn{2}{c|}{\textbf{LLaMA}}      & \textbf{Mistral}           & \textbf{Vicuna}          \\
\textbf{Datasets}      & \textbf{GPT-3.5} & \textbf{GPT-4o} & \textbf{LLaMA2-7b} & \textbf{LLaMA2-13b} & \textbf{Mistral-7b} & \textbf{Vicuna-7b} \\ \midrule
\textbf{Security News} & 0.880            & 0.962           & 0.895              &     0.918          & 0.900               &        0.883            \\
\textbf{Fiction}       & 0.895            & 0.985           & 0.875              &      0.899               & 0.940               &        0.865            \\
\textbf{IMDb}          & 0.862            & 0.904           & 0.820              &      0.811               & 0.855               &          0.794         \\  
\textbf{Medical}          &    0.861         &  -          &     0.811         &             0.827         &            0.801    &        0.775           \\ 
\bottomrule
\end{tabular}
}
\end{table*}

%-------------------------------------------------------------------------------
\section{Results}
\label{section: result}
%-------------------------------------------------------------------------------

%-------------------------------------------------------------------------------
\subsection{Evaluation Across Setups and Datasets}
\label{section: based results}
%-------------------------------------------------------------------------------

Our comprehensive evaluation of experimental settings and datasets reveals significant trends in the effectiveness of MIAs using the \attack{}. 
Across all configurations, we deploy consistent hyperparameters to ensure equitable comparison, such as the batch size.
The overall ROC\_AUC score indicates favorable results across various model setups and datasets.
The setup involves fine-tuning a target LLM using the datasets we collect and organize, then formulating specific queries to evaluate the LLM's understanding of the fine-tuned data.
For each dataset and LLM, we compute the average of the ROC\_AUC values from the four experimental results to obtain the final result.

The average ROC\_AUC scores for each LLM are illustrated in \autoref{figure:average}.
According to the results, the \attack{} demonstrates strong performance across all six target LLMs, with each LLM achieving a ROC\_AUC score of approximately 0.800 or exceeding.
The average ROC\_AUC score for the \attack{} is 0.873.
This indicates a clear distinction between member and non-member queries.
This performance remains consistently stable across multiple randomized trials, confirming the robustness of the \attack{}.

We also investigate the impact of various model architectures and evaluate our attack on four distinct LLM architectures.
Notably, GPT-4o, which boasts over 200 billion parameters~\cite{GPT4o}, achieves an average ROC\_AUC score of 0.950 in the three datasets.
Conversely, the mean performance of Vicuna-7b, which has 7 billion parameters~\cite{Vicuna}, is lower, exhibiting an average ROC\_AUC of merely 0.832 in the same datasets.
This indicates that GPT-4o's capacity for learning and understanding is superior among these models, making it more vulnerable to the \attack{}.
In other words, the \attack{} is most pronounced in GPT-4o.
Large parameter counts enhance the capability of LLMs as high-intelligence models, enabling them to perform complex and multi-step tasks more effectively.
In this regard, it is evident that an increase in the parameter size of LLMs corresponds with a higher likelihood of success for the \attack{}.

\begin{figure*}[!t]
\centering
\begin{subfigure}{0.5\columnwidth}
\includegraphics[width=\columnwidth]{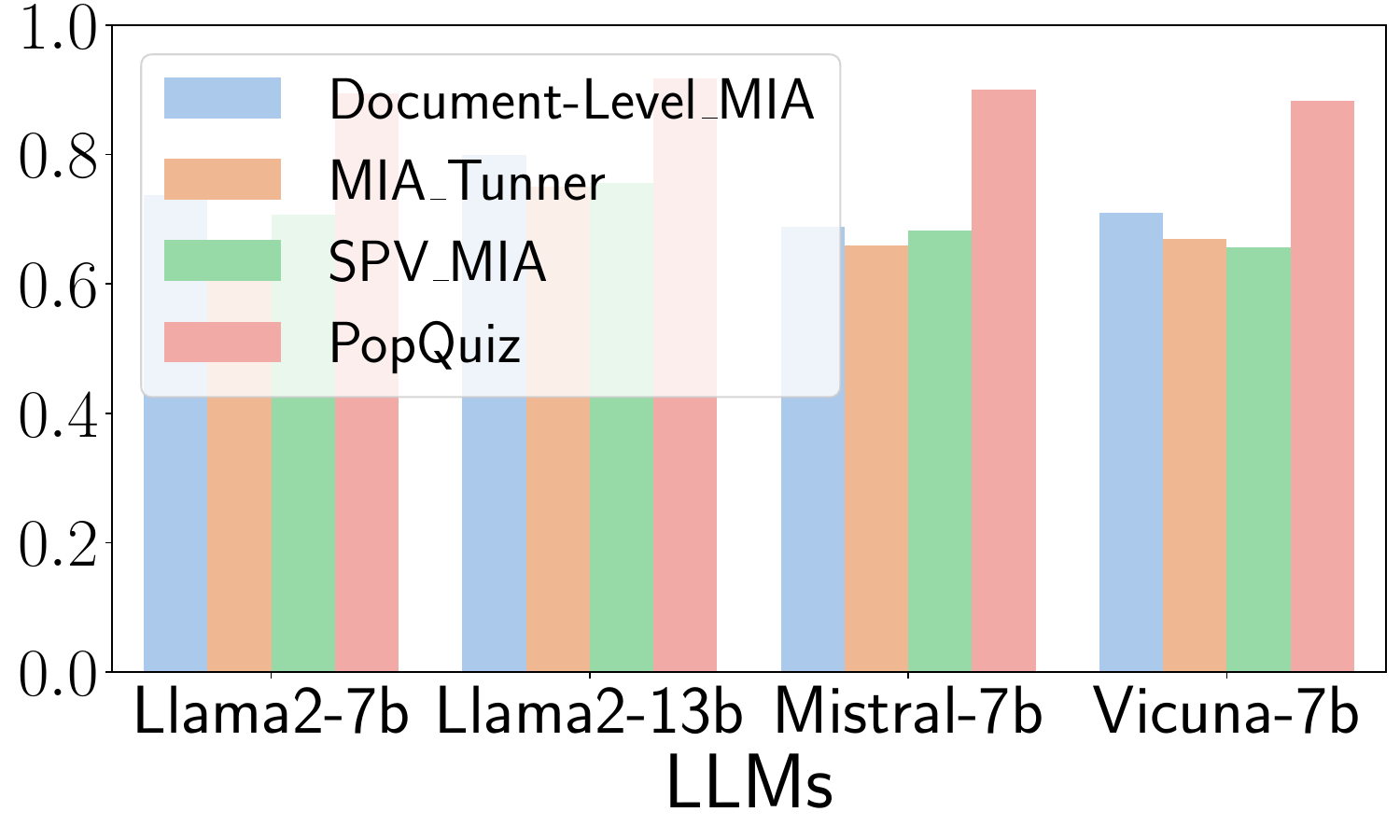}
\caption{Security News}
\label{figure:baseline a}
\end{subfigure}
\begin{subfigure}{0.5\columnwidth}
\includegraphics[width=\columnwidth]{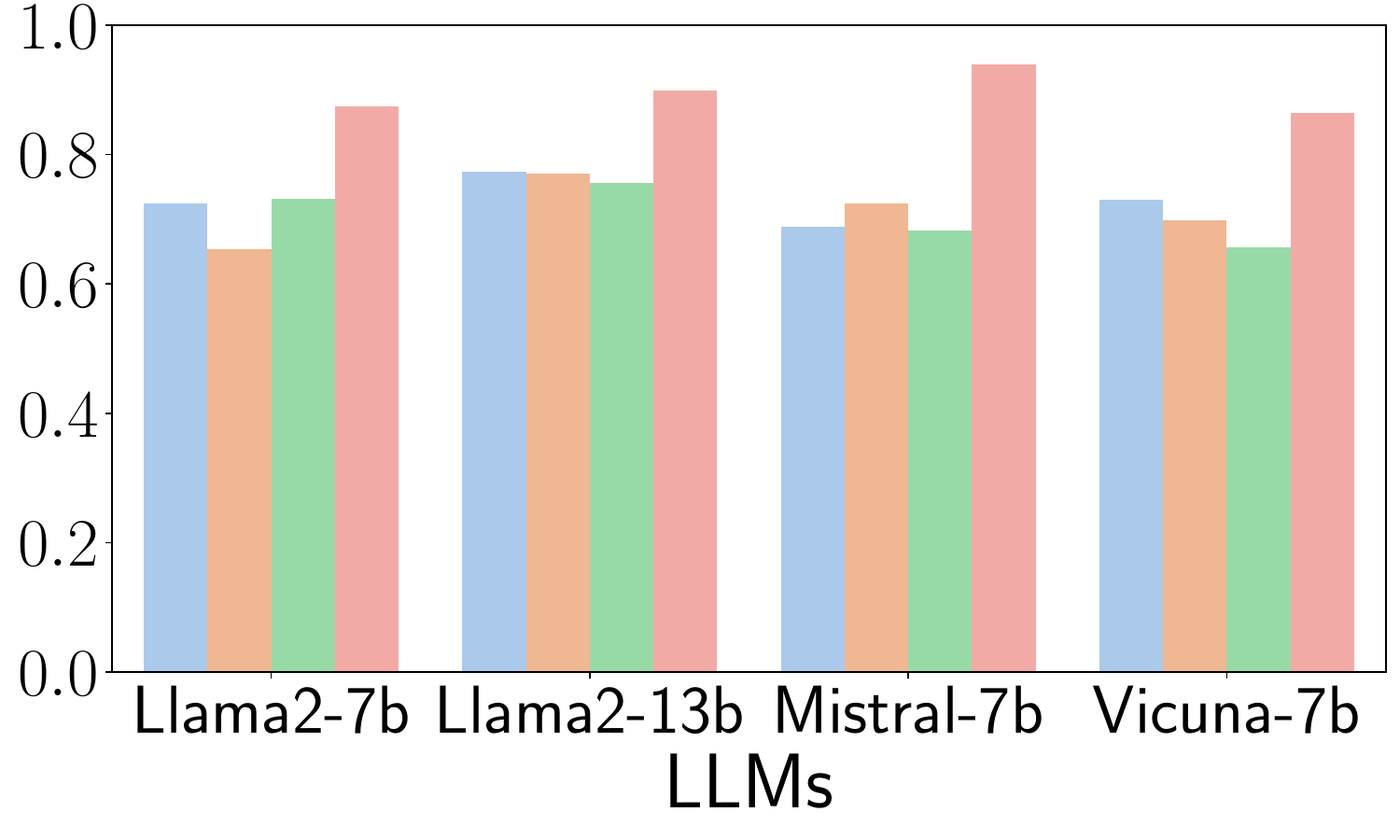}
\caption{Fiction}
\label{figure:baseline b}
\end{subfigure}
\begin{subfigure}{0.5\columnwidth}
\includegraphics[width=\columnwidth]{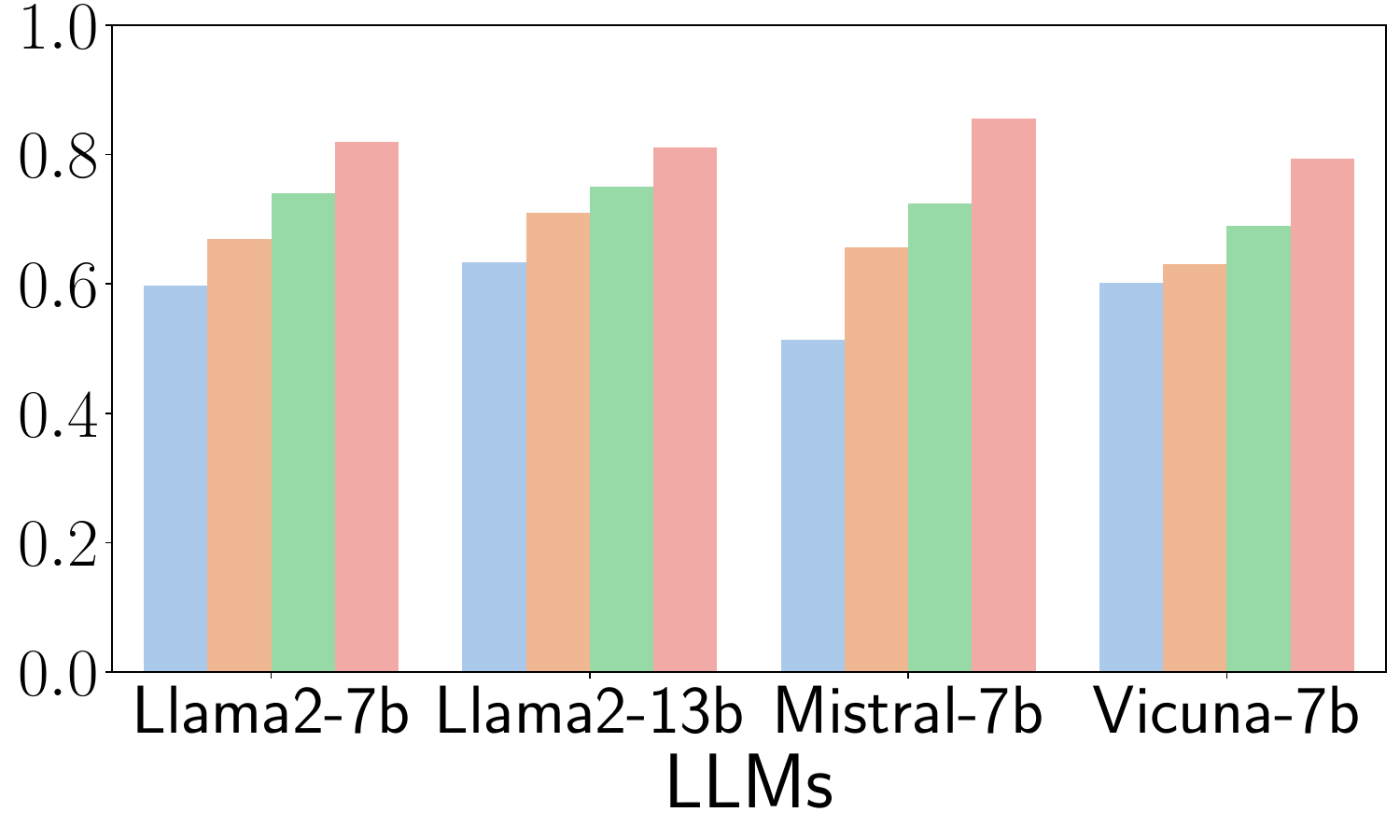}
\caption{IMDb}
\label{figure:baseline c}
\end{subfigure}
\begin{subfigure}{0.5\columnwidth}
\includegraphics[width=\columnwidth]{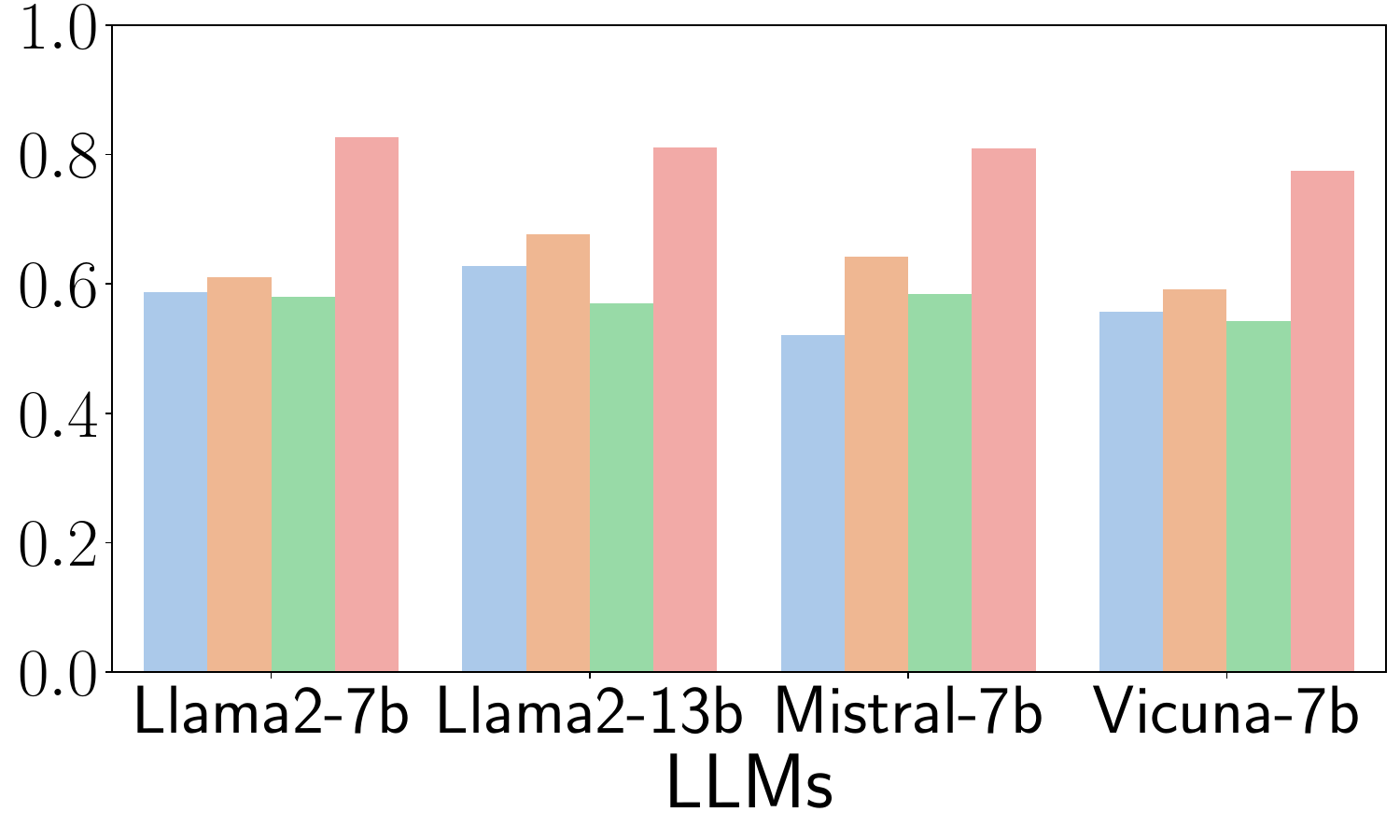}
\caption{Medical}
\label{figure:baseline d}
\end{subfigure}
\caption{Comparative ROC\_AUC scores show the baseline method performing best on \textit{Security News}, underperforming on \textit{IMDb}, and consistently demonstrating lower accuracy than \attack{} across all datasets.
}
\label{figure:baseline_compare}
\end{figure*}

As shown in \autoref{table:results} and \autoref{figure:log}, due to \textit{Medical} containing sensitive data and the LLM's security policy, the ROC\_AUC score is slightly lower than on other datasets.
Nevertheless, the \attack{} still achieves a high average success rate of 0.815 on \textit{Medical}. 
Also, \attack{}'s success rate varies inversely with dataset size, with smaller datasets providing more reliable attack surfaces. 
GPT-4o achieves the highest performance on the \textit{Fiction} dataset, with a ROC\_AUC score of 0.985, and on the \textit{IMDb} dataset, achieving a ROC\_AUC score of 0.904 with the same target LLM.
In contrast, the lowest performance on the \textit{IMDb} dataset is Vicuna-7b, which has a ROC\_AUC score of 0.794.
It can be seen that the large volume of data in \textit{IMDb} dataset has impacted the performance of the \attack{} to a certain degree.
In comparison to the \textit{IMDb} dataset, \textit{Fiction} and \textit{Security News} are smaller, and the \attack{} is least effective on the \textit{IMDb} dataset.
The \attack{} shows the best performance on the \textit{Fiction} dataset, achieving a mean ROC\_AUC score of 0.910, while it demonstrates high accuracy on the \textit{Security News} dataset, achieving a mean ROC\_AUC score of 0.906.
Simultaneously, it is evident that the \attack{} achieves the lower attack success accuracy on the \textit{IMDb} dataset, with an average ROC\_AUC score of 0.841.
The results show that none of the \attack{} achieve the same level of effectiveness on the \textit{IMDb} dataset as on the other two datasets, excluding the sensitive dataset \textit{Medical}. 
It may be because the other two datasets are more text-based.
As shown in \autoref{section:exp_dataset}, \textit{IMDb} dataset contains a significant amount of numerical data, which sets it apart from the other two datasets.
From this, we deduce that LLMs are less proficient in learning and understanding numerical-related data than in acquiring and comprehending text-related data. 
We examine and clarify this further in \autoref{section: diff_type}.

%-------------------------------------------------------------------------------
\subsection{Comparison with Baseline}
\label{section: baseline }
%-------------------------------------------------------------------------------
With the emergence of advanced LLMs,  there has been some recent research on performing MIAs against LLMs that can be employed as a baseline to compare with our performance~\cite{WLBZ24,MJRM24}.
To evaluate \attack{} under a unified protocol, we include three representative state-of-the-art MIAs that operate at the sequence/document level as follows:
\begin{itemize}[left=0pt]
    \item Document-Level MIA~\cite{MJRM24}. 
    It adopt a black-box document-level MIA that queries the target LLM for token-level predictions, normalizes these signals by token rarity, and aggregates them into document-level features.
    Then, a meta-classifier then predicts whether a document was included in the training data, and has been shown to generalize across model types and scales.
    \item MIA Tunner~\cite{FWGLLJ242}. 
    It formulates MIA as an instruction-based pre-training data detector by finetuning LLMs to directly answer whether a text was in their pre-training set. 
    This approach internally exploits the model's own responses to improve high-confidence detection across both aligned and unaligned models.
    \item SPV MIA~\cite{FWGLLJ24}. 
    It performs membership inference by constructing a self-prompt dataset from the target LLM itself to fine-tune a reference model and uses probabilistic variation grounded in the model's memorization behavior rather than overfitting to produce more reliable membership signals for fine-tuned LLMs.
\end{itemize}

Given that the baseline methods necessitate the application of a tokenizer in the experiment, we choose to use open-source LLMs for comparison.
Therefore, we use LLMs with tokenizers, namely LLaMA2-7b, LLaMA2-13b, Mistral-7b, and Vicuna-7b, across four datasets to run the baseline method.
The comparison with the baseline is shown in \autoref{figure:baseline_compare}.
We conduct the experiment using the open-source models, selecting the ROC\_AUC scores for comparison with our experiments.
From the results, we observe that the best-performing baseline, Document-Level MIA with LLaMA2-13b on \textit{Security News} achieves the highest performance when running target LLMs with the dataset utilizing the baseline with a ROC\_AUC score of 0.799.
However, even its best performance significantly lags behind the results achieved by the \attack{}.
Across all baseline methods, the baseline methods yield an average ROC\_AUC score of 0.667.
The average ROC\_AUC scores with \attack{} above 0.800 demonstrate that our method continues to perform well in various circumstances, indicating that \attack{} is more stable than the baseline methods.

Comparing our method to the baseline method reveals significant differences in performance, highlighting the advantages of \attack{}.
Moreover, our method outperforms the baseline with simpler, more natural queries, while the baseline requires complex query structures for effectiveness.

%-------------------------------------------------------------------------------
\subsection{Impact of Query Complexity}
\label{section: Query Complexity}
\begin{figure}
    \centering
    \includegraphics[width=0.80\linewidth]{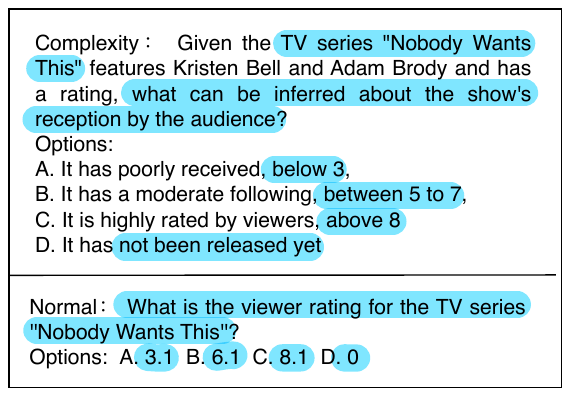}
    \caption{A comparative analysis example of query complexity, with highlighting indicating identical meanings across complex and normal (simple) queries.
    }
    \label{figure:example_query}
\end{figure}

%-------------------------------------------------------------------------------
In \autoref{section: Methodology}, the evaluation queries are designed to be simple and direct. 
To examine whether increased complexity affects attack performance, we construct more complex variants of the original \textit{IMDb} queries using GPT-4 while preserving their logical intent (\autoref{figure:example_query}). 
Detailed prompting strategies and examples are provided in ~\autoref{detail Query Complexity}. 
We evaluate these queries on fine-tuned LLaMA2-7b.
Contrary to expectations, complex queries reduce attack effectiveness. 
Simple queries achieve a ROC\_AUC score of 0.820, whereas complex queries obtain 0.720. 
The contextual information in complex queries does not enhance membership detection and may dilute factual recall signals. 
In contrast, direct and attribute-focused questions appear to better expose memorization behavior~\cite{KLHPLJK23,HRKSWH24}. 
We observe similar trends across different data types and categories, indicating that increasing query complexity does not improve attack performance.

%-------------------------------------------------------------------------------
\subsection{Impact of Different Data Types}
\label{section: diff_type}
%-------------------------------------------------------------------------------

\begin{table}[!t]
\centering
\caption{Results from separate inquiries reveal the \attack{} achieves significantly higher success rates on text-only data compared to number-only data.
}
\label{table:diff type}
\scalebox{0.90}{
\begin{tabular}{c|c}
\toprule
\textbf{Data Type} & \textbf{ROC\_AUC Score} \\
\midrule
\textbf{Text-Only}    &      0.880     \\
\textbf{Number-Only}  &      0.755  \\ 
\textbf{Mixed-type}  &       0.820   \\
\bottomrule
\end{tabular}
}
\end{table}

Since the \textit{IMDb} dataset contains substantial numerical fields (\autoref{section:exp_dataset}), we analyze how data type is affected \attack by comparing text-only, number-only, and mixed-type queries. 
Using the same query generation procedure (\autoref{section: Methodology}), we prompt GPT-4 to produce these three query sets for \textit{IMDb} and evaluate them on fine-tuned LLaMA2-7b.
As shown in \autoref{table:diff type}, text-only queries yield the strongest membership signals (ROC\_AUC 0.880), while number-only queries are less vulnerable (ROC\_AUC 0.755). 
This suggests that memorization of exact text values creates a larger attack surface. 
For number-only queries, shorter sequences (up to four digits) tend to be more vulnerable than longer ones; detailed breakdowns are provided in \autoref{detail data types}. 
Mixed-type queries achieve intermediate performance (ROC\_AUC 0.820), and numerical values embedded in text, like ratings, tend to be more vulnerable than purely numerical fields.

%-------------------------------------------------------------------------------
\subsection{Impact of Different Categories of Data}
\label{section: Different Categories}
%-------------------------------------------------------------------------------
\begin{table}[!t]
\centering
\caption{Performance results across datasets indicate highest effectiveness in the movie category, with mini TV series yielding the lowest performance metrics.
}
\label{table:diff categ}
\scalebox{0.90}{
\begin{tabular}{c|c}
\toprule
\textbf{Category} & \textbf{ROC\_AUC Score} \\
\midrule
\textbf{Moive}    &      0.850     \\
\textbf{TV mini Series}  &     0.770  \\ 
\textbf{TV Series}  &       0.810   \\
\bottomrule
\end{tabular}
}
\end{table}

Our analysis methodically explores distinct categories within the same dataset to fully understand their differing vulnerability to MIAs.
We analyze category-level differences in \textit{IMDb}, which includes movies, TV mini-series, and TV series (\autoref{detail experimental setup}). 
We sample 3,000 instances per category to form \textit{IMDb\_cate} and evaluate \attack{} on fine-tuned LLaMA2-7b; detailed dataset construction and evaluation setup are provided in \autoref{detail Different Categories}.
As shown in \autoref{table:diff categ}, movies are most vulnerable (ROC\_AUC 0.850), while TV mini-series yield the lowest score (ROC\_AUC 0.770), with TV series in between (ROC\_AUC 0.810). 
This pattern suggests that more established categories with richer and more consistent textual descriptions may produce stronger memorization signals. 
One possible explanation is that TV mini-series is a relatively recent category~\cite{Rise_of_the_Mini_Series}, and LLaMA2-7b’s knowledge cut-off may limit its familiarity with this category.

\begin{table*}[!t]
\centering
\caption{Evaluation results of defense mechanisms against the \attack{}.}
\label{table:defense}
\scalebox{0.80}{
\begin{tabular}{c|c|cc|cc|cc}
\toprule
     &                        & \multicolumn{2}{c|}{\textbf{GPT}}  & \multicolumn{2}{c|}{\textbf{LLaMA}}      & \textbf{Mistral}           & \textbf{Vicuna}          \\
\textbf{Defenses}                                   & \textbf{Dataset}       & \textbf{GPT-3.5} & \textbf{GPT-4o} & \textbf{LLaMA2-7b} & \textbf{LLaMA2-13b} & \textbf{Mistral-7b} & \textbf{Vicuna-7b} \\ \midrule
\multirow{4}{*}{\textbf{Instruction-Based Defense}} & \textbf{Security News} &  0.749   &  0.744   &  0.801       &    0.809   &   0.805    &   0.750    \\
     & \textbf{Fiction}   & 0.752  &  0.758     &  0.815   &   0.807    &   0.813      &   0.768     \\
     & \textbf{IMDb}  &  0.724    &  0.729 &     0.749  &    0.741  &  0.732    &   0.719  \\
     & \textbf{Medical}  &   0.709  &   -   & 0.753     &   0.727   &    0.712   &   0.680    \\ \midrule
\multirow{4}{*}{\textbf{Filter-Based Defense}}      & \textbf{Security News} &  0.811  &  0.795   & 0.727           &   0.739  &  0.717 &        0.722    \\
     & \textbf{Fiction} &  0.807   &  0.790    &            0.740 &   0.742 &   0.721     &  0.706   \\
     & \textbf{IMDb}   & 0.812    &  0.775   &  0.741  &     0.747  &  0.715    &     0.711      \\
     & \textbf{Medical}   &   0.722   &     -       &     0.635 &    0.657    &    0.665 & 0.677   \\ \midrule
\multirow{4}{*}{\textbf{DP-based Defense}}          & \textbf{Security News} &      0.745    &   0.751  &   0.723   &    0.738    &    0.730    &   0.725     \\
     & \textbf{Fiction}   &  0.720   &     0.725   &      0.702    &   0.735     &  0.718   &  0.715  \\
  & \textbf{IMDb}   &    0.695  &     0.710    &   0.728      &   0.733     &    0.707      &    0.716    \\
     & \textbf{Medical}   &    0.615    & -   & 0.700    &   0.717    &    0.655     &    0.620   \\ 
\bottomrule
\end{tabular}
}
\end{table*}
%-------------------------------------------------------------------------------
\subsection{Impact of Training Data Granularity}
\label{section: granularity}
%-------------------------------------------------------------------------------
We also examine how training data structure influences attack performance by comparing structured records with unstructured summaries derived from the \textit{Fiction} dataset. 
Across six LLMs, \attack consistently achieves higher ROC\_AUC scores on structured data than on unstructured data. 
For example, GPT-4o reaches 0.924 on the unstructured setting, while performance remains higher under the structured configuration. 
Overall, unstructured summaries reduce the strength of membership signals, suggesting that more organized and attribute-based training data increases vulnerability. 
Detailed experimental settings and full results are provided in \autoref{detail for granularity}.

%-------------------------------------------------------------------------------
\subsection{Impact of Different Learning Rates}
\label{section:result_diff_lr }
%-------------------------------------------------------------------------------
We further examine the effect of fine-tuning learning rates on attack performance using LLaMA2-7b and the \textit{Fiction} dataset. 
Across learning rates from 1e-4 to 4e-4, the ROC\_AUC scores vary within a narrow range (0.837 to 0.875). 
Although 3e-4 achieves the highest score, the overall differences remain small. 
These results indicate that \attack is not highly sensitive to moderate changes in fine-tuning learning rates. 
Detailed results are provided in \autoref{details for diff learning rates}.

%-------------------------------------------------------------------------------
\section{Defenses}
\label{section: defense}
%-------------------------------------------------------------------------------

The \attack{} effectively predicts the membership status of target samples, indicating significant privacy vulnerabilities. 
Current understanding lacks comprehensive defense methods against LLM-targeted MIAs~\cite{WLBZ24}.
We examine three defenses designed to reduce information leakage from LLMs regarding their prompts.

\mypara{Instruction-Based Defense}
It is a method of safeguarding LLMs by embedding carefully crafted directives within prompts to proactively steer the model away from generating harmful or undesirable outputs~\cite{ZLWJZBSZ24}.
Implementation details are provided in \autoref{detail defense}.

\mypara{Filter-Based Defense} 
It is a protective strategy in large language models that detects and blocks unsafe or malicious inputs or outputs using external filtering mechanisms~\cite{WLBZ24}.
Implementation details are provided in \autoref{detail defense}.

\mypara{DP-Based Defense}
Differential privacy (DP) has been recognized as a crucial protective measure against MIAs. 
DP-based defense in LLMs is a method that integrates differential privacy into the training process to prevent the exposure of sensitive data and protect the anonymity of individual records~\cite{TSIMMLGKS23}.
Implementation details are provided in \autoref{detail defense}.

\mypara{Results}
As shown in \autoref{table:defense}, \attack{} achieves an average ROC\_AUC of 0.873 without any protection, consistently exceeding 0.800 across most target LLMs and datasets. 
All three defenses reduce attack effectiveness to varying degrees. 
Instruction-based and filter-based defenses lower the mean ROC\_AUC to 0.753 and 0.731, respectively, indicating moderate mitigation. 
DP-based defense achieves the largest reduction, with an average ROC\_AUC of 0.708. 
Despite these reductions, attack performance remains substantially above random guessing across models, with ROC\_AUC scores frequently exceeding 0.700. 
Notably, GPT-4o exhibits particularly high vulnerability in the unprotected setting (ROC\_AUC $>$ 0.900), and although defensive measures reduce this value into the 0.700–0.800 range, membership signals remain detectable. 
These results suggest that existing defenses can weaken but do not eliminate membership leakage in LLMs.

%-------------------------------------------------------------------------------
\section{Conclusion}
\label{section: Conclusion}
%-------------------------------------------------------------------------------
In this study, we propose \attack{}, a black-box membership inference attack that evaluates memorization in LLMs through quiz-style queries.
By framing membership inference as a factual recall assessment, our method provides a simple yet effective way to expose training data signals from model outputs alone. 
Across all six models and four datasets, \attack{} achieves strong performance and improves over prior methods by 20.6\% on average, highlighting persistent privacy risks even under existing defenses. 
Although defensive strategies reduce attack success rates, they fail to eliminate membership leakage entirely.
These findings motivate stronger privacy-preserving training and deployment strategies for LLMs.

\section*{Ethical Consideration}
In this research, we examine MIAs targeting LLMs, which raise important ethical concerns about privacy and data security.
To stay ethically compliant, we strictly use publicly available datasets and models, ensuring that no proprietary or sensitive information is exposed.
Our experiments are solely intended to analyze LLM vulnerabilities to encourage the development of more robust privacy-preserving measures.
No private or confidential datasets were accessed.
Since our study does not involve human subjects or personal data, it does not require Institutional Review Board (IRB) approval.
We believe in transparency and responsible disclosure, sharing our findings with the community to strengthen LLM robustness.
All research artifacts, including datasets and code, will be shared under strict guidelines to prevent unethical exploitation.
While upholding the highest ethical standards, we hope to increase awareness of LLM privacy problems and assist in developing safer LLM systems.

\section*{Limitations}
Nonetheless, our research also revealed several limitations that warrant discussion. 
The effectiveness of the \attack{} varies significantly depending on the scale and diversity of the training datasets.
In cases where the training dataset is particularly small, the target LLM may leak sufficient distinct behavior patterns to enable dependable membership inference.
On the other hand, a large training dataset results in broader model behavior, making the identification more challenging of individual data membership with high confidence.
Another limitation of the \attack{} lies in its computational demands.
Although the attack mechanism is straightforward, generating effective queries requires careful consideration and multiple iterations.
While our current results indicate favorable attack success rates against standard techniques, the efficacy against models trained with specific privacy-preserving methods requires further investigation.

Future work presents several intriguing directions based on our findings.
By integrating natural language understanding and generation techniques, we can formulate more targeted and effective queries to enhance the \attack{} against LLMs. 
Examining utility-privacy trade-offs in LLMs can provide crucial insights for implementations involving sensitive data.
Techniques such as differential privacy, adversarial examples, or other privacy-preserving adaptation methods may mitigate the risks identified by our attack.

%-------------------------------------------------------------------------------
\bibliographystyle{plain}
\bibliography{normal_generated_py3}
%-------------------------------------------------------------------------------

\appendix

\section{More Details for Experimental Setup}
\label{detail experimental setup}
%-------------------------------------------------------------------------------
\subsection{Target Models}
\label{section:exp_models}
%-------------------------------------------------------------------------------
We consider six predominant LLMs to ensure thorough examination and verification of the efficacy of our MIA methodology, comprising both open-source and closed-source options. 
The specific information about the LLMs we have chosen is the following:
\begin{itemize}[left=0pt]
    \item \textbf{Closed-source LLMs.} 
    We include GPT-3.5-turbo-0125~\cite{GPT-3.5-Turbo}, employing the transformer architecture for enhanced text generation and GPT-4o-2024-08-06~\cite{GPT4o}, OpenAI's efficient model for complex NLP tasks with double speed at half the cost.
    \item \textbf{Open-source LLMs.} 
    We include LLaMA-2-7b-chat-hf~\cite{LLaMA2-7B-Chat} and LLaMA2-13b-chat-hf~\cite{llama2_13b_chat} from Meta AI; Mistral-7b-Instruct-v0.2~\cite{Mistral-7B-Instruct-v0.2}, a compact high-performance model; and Vicuna-7b-v1.5~\cite{Vicuna}, a LLaMA-based model optimized for dialogue tasks.
\end{itemize}

We use GPT-4o and GPT-3.5 as representatives of closed-source LLMs, conducting experiments via the OpenAI API~\cite{OpenAIAPI} with their default configurations.
For open-source LLMs, we use LLaMA2-7b, LLaMA2-13b, Mistral-7b, and Vicuna-7b as they balance computational feasibility and performance efficiency.
We access the models from Hugging Face, maintaining default configurations while conducting additional ablation studies on hyperparameters with varying learning rates.

\begin{table}[!t]
\centering
\caption{Knowledge cut-off dates of LLMs.
}
\label{table:models}
\scalebox{0.80}{
\begin{tabular}{c|cc}
\toprule
\textbf{Arch} & \textbf{Target LLM} & \textbf{Knowledge Cut-off Date} \\
\midrule
GPT       & GPT-4o-2024-08-06     & Oct. 2023              \\
  & GPT-3.5-turbo-0125 & Sep. 2021 \\ \midrule
LLaMA  & LLaMA2-7b-chat-hf   & July. 2023                     \\
  & LLaMA2-13b-chat-hf & July. 2023  \\ \midrule
Mistral    & Mistral-7b-Instruct-v0.2    &   Sep. 2023           \\ \midrule
 LMSYS  &  Vicuna-7b-v1.5 & March. 2023 \\
\bottomrule
\end{tabular}
}
\end{table}

%-------------------------------------------------------------------------------
\subsection{Datasets}
\label{section:exp_dataset}
%-------------------------------------------------------------------------------

LLMs train on diverse public data sources to enhance performance and generalization~\cite{LCLDJ24}. 
To evaluate our attack, we partition the datasets into two segments: one half as members and another half as non-members. 
Although some of the open-source models have disclosed partial LLM configurations and training insights, precise training datasets remain unavailable. 
Given the lack of ground truth membership status, we collect new datasets that are published after the knowledge cutoff dates of all target LLMs and fine-tune subsets into target LLMs as members, ensuring both member and non-member data maintain similar distributions~\cite{FKYTB24,LCLDJ24}.
To find suitable fine-tuning datasets, we collected the official knowledge cut-off dates of all target models and incorporated the information in \autoref{table:models}.

\begin{table}[!t]
\centering
\caption{Statistics of the datasets.}
\label{table:dataset-related}
\scalebox{0.80}{
\begin{tabular}{c|cc}
\toprule
\textbf{Dataset} & \textbf{\# Samples} & \textbf{Fields} \\
\midrule
 Security News    &      1,350  &  Title, Date, Author,     \\
 & &  Category, Keywords      \\ \midrule
 Fictions  &       864 &   Title, Published Country, \\ 
 &  & Status, Category, Chapter             \\ \midrule
  IMDb  &       10,000  &   Title, Type, Intro, Vote, \\
  & & Rate, Certificate, Category      \\\midrule
    Medical  &       5,000  &   Name, Age, Hometown, BMI,  \\
  & & Gender, Treatment, Bloods, Hospital     \\
\bottomrule
\end{tabular}
}
\end{table}

Based on \autoref{table:models}, the datasets collected for experiments should be published after the cut-off dates, specifically in January 2024 or later.
To facilitate the smooth implementation of our proposal, the dataset selected for the experiments excludes toxic, illegal, and other harmful information. 
Due to human rights and ethics review board (IRB) constraints, sensitive datasets are difficult to access, necessitating our use of open-source data in experiments.
Most MIA studies similarly utilize open-source datasets~\cite{XWHZGPGD24,MJRM24,CTCP21,CCNSTT22,CSSWZ22,YMMBS22,WLBZ24,MJRM24,CYZF20}. 
Nevertheless, we choose one synthetic dataset, \textit{Medical} dataset, containing sensitive information among numerous open-source resources.
Finally, we collect four datasets, namely Security News, Fiction, IMDb, Medical, in our experiments.
Each dataset is split equally into members and non-members for LLM fine-tuning and evaluation.
Dataset-related information is shown in \autoref{table:dataset-related}.
The statistics of the datasets is shown in \autoref{table:dataset-related}

\noindent\textbf{Securelist~\cite{securelist}} is from Kaspersky Lab's cybersecurity platform. It provides insights on cyber threats, vulnerabilities, and security trends, covering threat research, malware analysis, digital privacy, and emerging threat intelligence.

Concretely, we crawl and collect 1,350 webpages from Securelist, ranging from January 2024 to September 2024.
For each webpage, we extract multiple components, including the news title, publication date, author, category, related keywords, etc.
For specific use in the experiment, we further filter the dataset by focusing on components that include easily verifiable factual statements, such as publication date and author.
We implement data cleaning procedures, removing special characters, standardizing formatting, and ensuring consistent capitalization and punctuation. 
After cleaning, we extract relevant fields, like title, author, date, category, eliminating unhelpful symbols and hyperlinks to facilitate paragraph integration. 
This preprocessing methodology is consistently applied across all datasets to ensure reproducibility.

After organizing the dataset, we merge each data component into a relevant story paragraph, which assisted in generating question-answer pairs for our membership inference attacks and fine-tuning the target LLM. 
Examples of the processed data samples can be found in \autoref{securelist}.
The format is:
\begin{quote}
    \noindent
    \textbf{``\textit{[News Title]} is posted on \textit{[Publication Date]}, and it is written by \textit{[Author]}.
    The category for \textit{[News Title]} is \textit{[Category]}, and the keywords for it are \textit{[Keywords]}.''}
\end{quote}

\noindent\textbf{Fiction~\cite{Fantasy_Manga_Datasets}} is collected by Prem Mevada from Kaggle. 
It is a vast compilation of metadata containing numerous manga series, including a detailed summary of their characteristics and present conditions.

The dataset includes 865 data points, summarizing manga information from January 2024 until the present.
This dataset is optimal for analyzing, studying, and understanding trends in the manga industry.
We remove a random piece of data to make the data set double-digit, 864 data points in total, for experiments.
Due to extensive redundant data, like titles, subtitles, authors, and summaries, preprocessing is required to distill key findings from each data point. 
Language standardization is necessary with multiple languages present, including English, Chinese, and Japanese. 
We removed multilingual \textit{author} and \textit{subtitle} categories to unify the dataset to English.
Following that, we review irrelevant categories that are unrelated to the data points and are tailored by the dataset developer, like data IDs and data upload IDs. 
We remove the unnecessary categories from the dataset, resulting in a refined dataset suitable for direct use.

We adhere to the prior methodology of combining each data component within the structured and accessible dataset into pertinent text paragraphs for the experiments.
Processed data samples are provided in \autoref{fiction}.
The format is: 
\begin{quote}
    \noindent
    \textbf{``The manga named \textit{[Title]} is from \textit{[Published Country]}, and its status is \textit{[Status]}.
    The category of \textit{[Title]} is \textit{[Category]}. 
    The total number of chapters for \textit{[Title]} is \textit{[Chapter]}.''}
\end{quote}

\noindent\textbf{IMDb~\cite{IMDB}} is the premier online repository for film, television, and entertainment industry data. 
Established in 1990 and acquired by Amazon, it houses millions of titles and profiles, providing filmographies, cast information, synopses, user ratings, reviews, and industry news.

We gather and crawl IMDb record data from January 2024 to September 2024, resulting in a dataset that includes 64,834 data points. 
This dataset can be preprocessed to extract essential metrics and statements for IMDb records.
This dataset comprises thorough metadata for each data point, covering release dates, ratings, and the total number of ratings, which can provide insights into the overall success of items within the dataset.
We randomly selected 10,000 samples from a total of 64,834 for the experiment.
Initially, we preprocess the crawled data by removing irrelevant information, such as poster size, Metascore, poster number, etc., from the dataset.
Finally, we obtain clean data suitable for experiments.

We follow the previous method of integrating each data component within the accessible dataset into relevant text paragraphs.
Sample processed data is available in \autoref{imdb}.
The structure is: 
\begin{quote}
    \noindent
    \textbf{``The type of \textit{[Title]} is \textit{[Type]}. 
    The introduction to \textit{[Title]} is \textit{[Intro]}. 
    The certificate of \textit{[Title]} is \textit{[Certificate]} and the category is \textit{[Category]}.
    \textit{[Vote]} people voted for \textit{[Title]}, and the rating is \textit{[Rate]}.''}
\end{quote}

\noindent\textbf{Medical~\cite{Heart_Disease_Dataset}} is a Kaggle dataset collected by Saquib Hazari, comprising synthetic patient records specifically created for heart disease research, incorporating disease-relevant characteristics.

The dataset includes 10,000 data points and is published in October 2024.
It summarizes patient information relevant to various aspects of heart disease diagnosis and prediction.
It establishes an essential domain for assessing predictive analytics and machine learning models focused on cardiac health, but poses a potential risk of data leakage from itself.
During the experiments, we randomly select 5,000 data points from the total. 
The collection contains extensive specialist information pertaining to heart disease, and it is well organized.
Consequently, we evaluated this particular data, such as cholesterol, in our tests.
Ultimately, only two specialized data points, BMI and blood pressure, remained to be included in the experiment alongside other patient information.

We keep to the prior methodology of incorporating each data element within the readily accessible dataset into pertinent textual paragraphs.
\autoref{medical} contains examples of the processed data.
The setup is as follows: 
\begin{quote}
    \noindent
     \textbf{``\textit{[Name]}is living in \textit{[hometown]}, USA. 
     The age of \textit{[Name]} is \textit{[Age]}.
     The gender of \textit{[Name]} is \textit{[Gender]}.
     \textit{[Name]} has treated in \textit{[Hospital]}. 
     The treatment for \textit{[Name]} is \textit{[Treatment]}. 
     \textit{[Name]}'s blood pressure is \textit{[Blood\_pressure]}, and \textit{[Name]}'s BMI is \textit{[BMI]}.''}
\end{quote}

%-------------------------------------------------------------------------------
\subsection{Training and Evaluation Setup}
\label{section:exp_eval}
%-------------------------------------------------------------------------------

Our training and evaluation workflow is structured to meticulously evaluate our MIA method while guaranteeing reproducibility and statistical validity.

As mentioned in \autoref{section:exp_dataset}, the primary setup involves fine-tuning a target LLM with our collected datasets, followed by crafting specific queries designed to assess the LLM's comprehension of the training data.
The implementation utilizes the Transformers library from Hugging Face, along with their PEFT toolkit, for effectively fine-tuning target LLMs.
We divide the dataset into members and non-members to train and validate our method for predicting dataset membership within fine-tuned datasets.
Each of our gathered datasets is equally split into members and nonmembers.
Next, we input the organized data and apply the GPT-4 API to construct four multiple-choice questions for each data sample according to the specified prompt.
Finally, we gather all responses provided by the fine-tuned target LLMs, and extract the answer options it gives. 

To determine whether a data point is a member or not, we compare the responses given by the fine-tuned target LLM with the correct answers for each question.
The data assigned for members is adjusted to satisfy the PEFT requirements and is fine-tuned into target LLMs.
For the evaluation, the primary metric is the ROC\_AUC score for attack effectiveness.
All trials are conducted three times to ensure statistical reliability. 
We employ bootstrapping to compute confidence intervals and perform both automated and manual validation of results.
This evaluation framework guarantees comprehensive validation of our attack approach while offering in-depth insight into the factors that affect attack success rates.

For the fine-tuning of GPT-4o and GPT-3.5, knowing that they are closed-source LLMs, we leverage OpenAI's functionality to automatically select fine-tuning parameters according to the dataset size.
We employ LoRA as the fine-tuning method for the open-source LLMs, LLaMA2-7b,  LLaMA2-13b, Vicuna-7b, and Mistral-7b.
We modify the learning rate to 3e-4, set the batch size to 128, and establish the number of epochs for each dataset during fine-tuning at 100.
During the fine-tuning of Vicuna-7b, we observe that when the epochs are at 100, the loss fails to converge effectively, and the loss ends up converging at around 2.87 at 100 epochs.
We increase the epochs to 200 for fine-tuning Vicuna-7b; we see that the loss coverage is better, and the loss ends up around 3e-4.
Therefore, we modify the epochs for Vicuna-7b to 200 for experimental purposes.
For LLaMA2-13b, due to the excessive number of its parameters, we adjust the batch size to 64, while maintaining consistency with the configuration of other open-source LLMs.
Upon conforming to this parameter configuration, the fine-tuning of the LLM can be successfully carried out.

%------------------------------------------------------------------------------- 
\section{More Details for Impact of Query Complexity} 
\label{detail Query Complexity} 
%-------------------------------------------------------------------------------
The queries previously created for \textit{IMDb} dataset are made more complex using GPT-4.
We instruct GPT-4 to stay true to the logic of the normal (simple) queries to enhance their complexity.
To generate complex query variants, we prompt GPT-4 using the following template:
\begin{quote}
Generate a multiple-choice question that tests a factual attribute of this record. 
The question should be context-rich and syntactically more complex than a direct factual question. 
Do not introduce new facts beyond the provided record. 
Ensure that the correct answer is uniquely determined by the given data.

Data: [Data]

Provide: \\
- A multiple-choice question \\
- Four answer options \\
- Clearly indicate the correct answer
\end{quote}
We ensure that the correct answer remains unchanged after rewriting and manually verify a random subset of generated queries for consistency.

Abundant information within complex queries fails to improve the target fine-tuned LLM's comprehension or its ability to respond accurately.
These complex queries prevent the target fine-tuned LLM from integrating factual recall with contextual understanding.
Conversely, since normal (simple) queries are posed in a direct and clear manner, they facilitate better responses from the target fine-tuned LLM~\cite{KLHPLJK23,HRKSWH24}.

%-------------------------------------------------------------------------------
\section{More Details for Impact of Different Data Types}
\label{detail data types}
%-------------------------------------------------------------------------------
The results are displayed in \autoref{table:diff type}.
Text-only data consistently exhibits a higher vulnerability to MIAs, achieving a ROC\_AUC score of  0.880, whereas number-only data shows more variable results with a ROC\_AUC score of 0.755.
In the context of text-only data, we observe particularly high attack success rates when targeting words with elevated precision. 
Textual data sequences reveal the greatest vulnerability, highlighting that the target LLM's retention of exact text values generates a significant attack surface.
Number-only data primarily reveals more complex patterns in the attack success rates when dealing with floating-point data.
Short numerical sequences, consisting of one to four digits, display relatively high vulnerability, with the model correctly answering at least half of the four multiple-choice questions.
In contrast, brief numerical sequences ($<$4 digits) offer more distinctive patterns for membership inference, while more extended sequences show lower vulnerability with models answering at most two of four questions correctly.
The combination of numerical and textual data within training instances exhibits unique characteristics.
Mixed-type queries indicate attack success rates that are intermediate between those of number-only and text-only data, with a ROC\_AUC score of 0.820.
Observation of number-only data incorporated within the text, such as movie ratings, indicates higher vulnerability compared to number-only data.

%-------------------------------------------------------------------------------
\section{More Details for Impact of Different Categories of Data}
\label{detail Different Categories}
%-------------------------------------------------------------------------------
\autoref{section:exp_dataset} clarifies that the crawled \textit{IMDb} dataset comprises three categories: movies, mini TV series, and TV series.
From the original \textit{IMDb} dataset, which contains 64,834 data points, we randomly select 3,000 samples from each of the three categories: movies, mini TV series, and TV series, resulting in a new dataset containing 9,000 samples, referred to as \textit{IMDb\_cate}.
With the \textit{IMDb\_cate} dataset evenly distributed among the three categories, we note considerable variation in the attack success rate.
The dataset is divided into 4,500 member samples and 4,500 non-member samples in a ratio of 1:1.
This ensures that there are 1,500 movie data points, 1,500 mini TV series data points, and 1,500 TV series data points for both members and non-members.
The sample of 4,500 members is fine-tuned into a target LLM, specifically the LLaMA2-7b used in this case.
Next, a set of queries derived from 9,000 data points, each containing four related multiple-choice questions, is input into the fine-tuned LLaMA2-7b for evaluation to obtain the final results.

%-------------------------------------------------------------------------------
\section{More Details for Impact of Different Learning Rates}
\label{details for diff learning rates}
%-------------------------------------------------------------------------------

\begin{table}[!t]
\centering
\caption{Results for learning rates on LLaMA2-7b using \textit{Fiction} dataset.
}
\label{table:diff lr}
\scalebox{0.85}{
\begin{tabular}{c|c}
\toprule
\textbf{Learning Rate} & \textbf{ROC\_AUC Score}  \\
\midrule
 \textbf{1e-4}   &      0.837       \\
 \textbf{3e-4}  &      0.875  \\ 
\textbf{4e-4}  &       0.855   \\
\bottomrule
\end{tabular}
}
\end{table}

We utilize the LLaMA2-7b model with the \textit{Fiction} dataset to examine the impact of fine-tuning with different learning rates.
Our investigation explores learning rates of 1e-4, 3e-4, and 4e-4, observing distinct patterns in attack effectiveness.
Specifically, we examine the learning rates 1e-4, 3e-4, and 4e-4.
Upon reviewing the loss after fine-tuning the target LLM, we observe minor fluctuations in the loss during the fine-tuning phase at a learning rate of 4e-4.
Ultimately, the loss of all three fine-tuned target LLMs converges to a range between 0.0002 and 0.0032, indicating the efficacy of the fine-tuning process.
Finally, we evaluate each of the three fine-tuned target LLMs using the query set associated with the \textit{Fiction} dataset.
Learning rates are an important parameter during LLM fine-tuning.
To investigate their impact, we take LLaMA2-7b as a case study.
Specifically, we first fine-tune LLaMA2-7b with \textit{Fiction} dataset using three learning rates: 1e-4, 3e-4, and 4e-4.
We then perform the \attack{} on each of the three fine-tuned LLMs.
The results for the three different learning rates are illustrated in \autoref{table:diff lr}.

Higher learning rates of 3e-4 and 4e-4 produce stronger membership signals, attaining a ROC\_AUC score of up to 0.850.
Lower learning rates of 1e-4 yield less stronger membership signals, with a ROC\_AUC score of 0.837.
These results indicate that a learning rate of 3e-4 achieves optimal performance in our experiments, with a ROC\_AUC score of 0.875.
In other words, we identify 3e-4 as the optimal learning rate, at which the \attack{} demonstrates a high success rate while maintaining stability across various configurations compared to the other two learning rates.
Nonetheless, in terms of the ROC\_AUC score of the \attack{}, the impact of the attack remains largely consistent.
Thus, modifying the learning rate within the range of 1e-4 to 4e-4 during the experiment has minimal impact on our attack methodology.

%-------------------------------------------------------------------------------
\section{More Details for Impact of Training Data Granularity}
\label{detail for granularity}
%-------------------------------------------------------------------------------
\begin{figure}[!t]
\centering
\begin{subfigure}{0.45\columnwidth}
\includegraphics[width=\columnwidth]{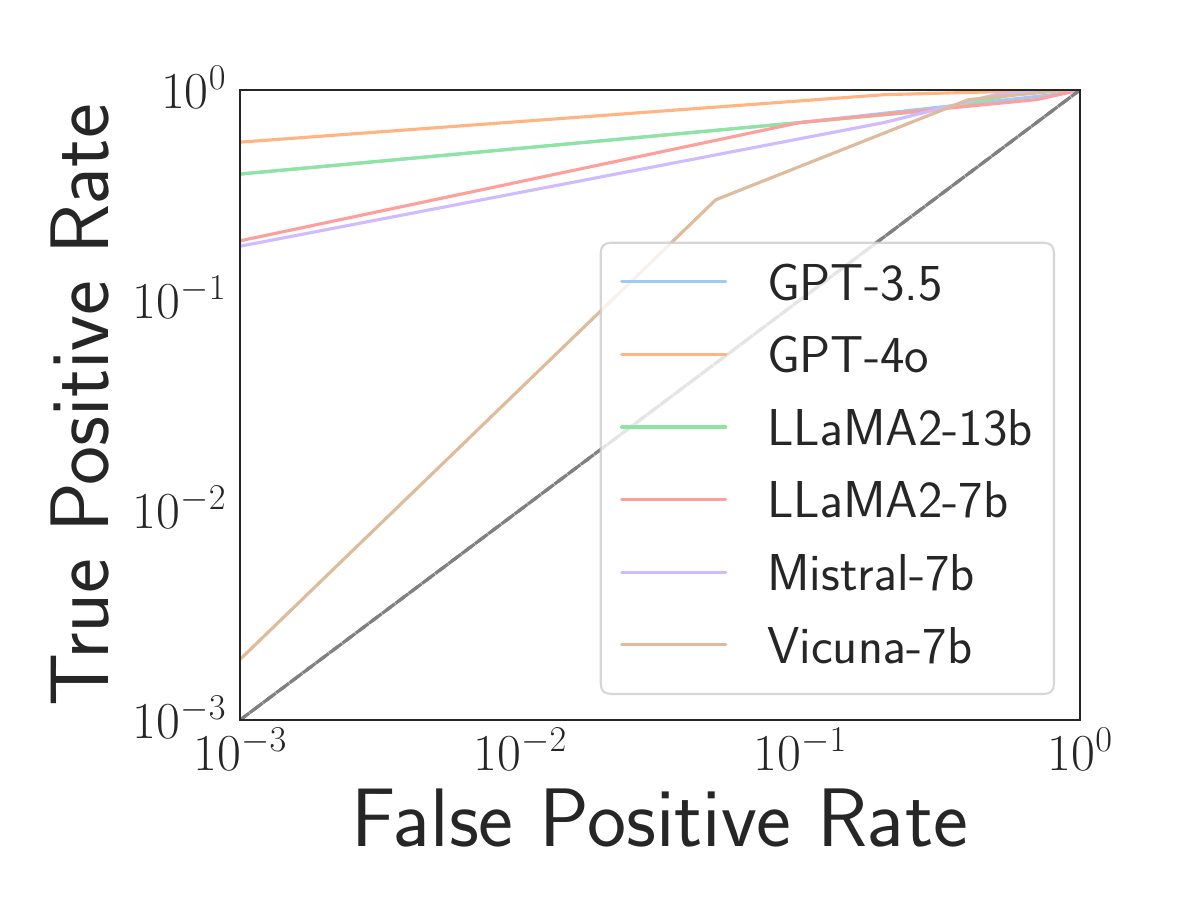}
\caption{Log-scale ROC Curve}
\label{figure:abs log}
\end{subfigure}
\begin{subfigure}{0.45\columnwidth}
\includegraphics[width=\columnwidth]{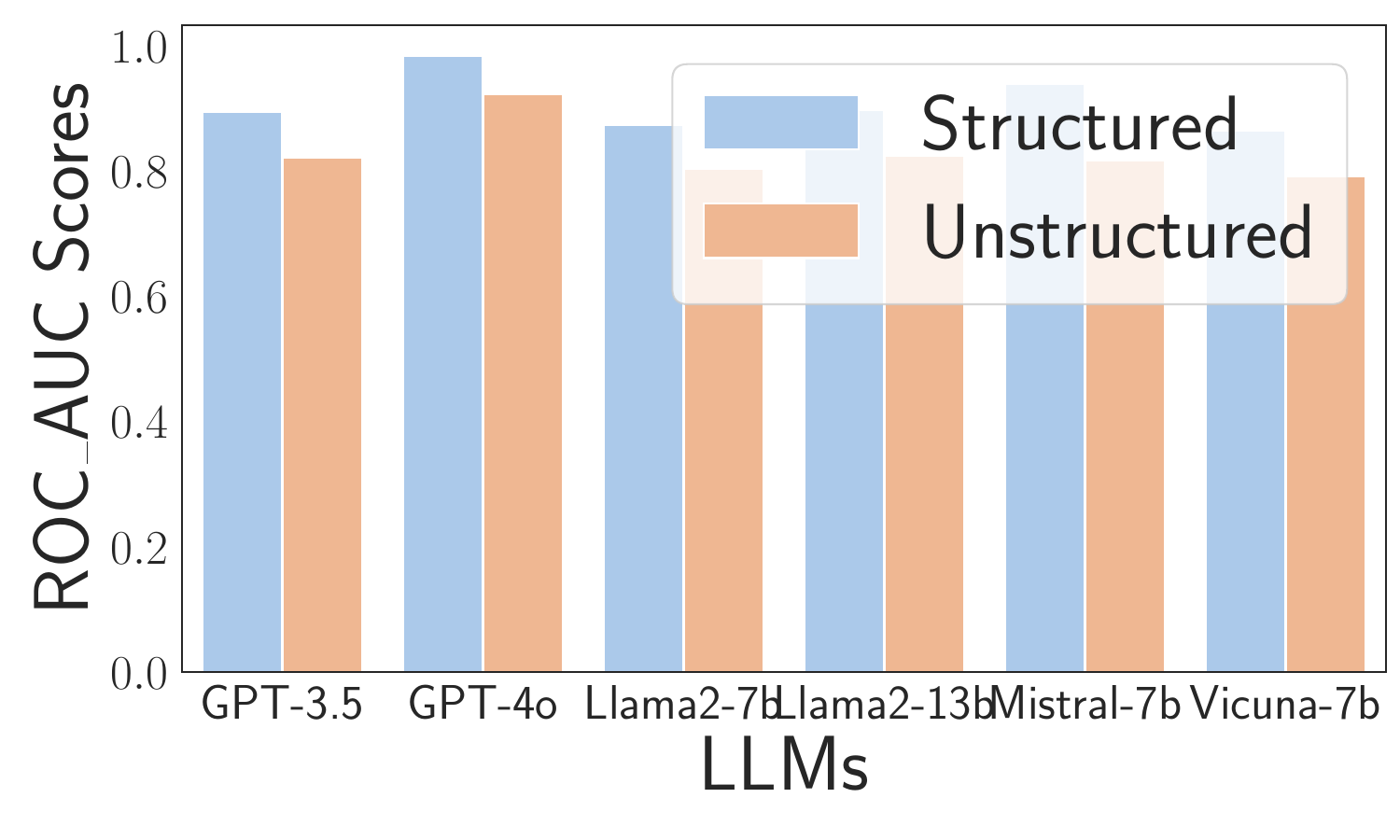}
\caption{Structured vs. Unstructured}
\label{figure:abs comp}
\end{subfigure}
\caption{The \attack{} achieves higher ROC AUC values with structured versus unstructured data across various LLM architectures, with GPT-4o exhibiting the greatest vulnerability regardless of data type.
}
\label{figure: abd}
\end{figure}

In this section, our examination of input granularity uncovers significant insights regarding the impact of training data architecture on the \attack{} success rate.
We implement one approach, using a summary for each data point from \textit{Fiction} dataset as original data, maintaining natural document flow.
We designate the data processed in this manner as ``unstructured'' data and refer to \textit{Fiction\_abs} dataset.
Examples of data in the \textit{Fiction\_abs} dataset can be found in \autoref{fiction_abs}.
With \textit{Fiction\_abs} dataset, separating the 864 summary-only data points into members and non-members in a 1:1 ratio.
We input 432 member data into the six target LLMs utilizing fine-tuning.
Following that, four multiple-choice questions are generated for each data point in \textit{Fiction\_abs} individually.
Then, those queries are incorporated into the fine-tuned target LLMs for the \attack{} evaluation.
The result is shown in \autoref{figure:abs log}.

The \attack{} also targets \textit{Fiction\_abs} dataset, an unstructured dataset.
The \attack{} reaches the highest ROC\_AUC score with GPT-4o, 0.924, demonstrating superior vulnerability as evidenced by its elevated curve trajectory.
Vicuna-7b has the lowest ROC\_AUC score, 0.793.
This finding aligns with our conclusion in \autoref{section: based results}.
It is once again demonstrated that the \attack{} outperforms on GPT-4o while underperforming on Vicuna.
We also compare it with the result we get in \autoref{section: based results}.
For the purpose of comparison, we donate the data utilized in the experimental setup as ``structured'' data. 
\autoref{figure:abs comp} shows that the \attack{} generally achieves higher ROC\_AUC scores when using structured data compared to unstructured data across six target LLMs.
The results repeatedly indicate that unstructured data significantly reduces vulnerability to membership inference attacks in comparison to structured data.

%-------------------------------------------------------------------------------
\section{More Detail for Defenses}
\label{detail defense}
%-------------------------------------------------------------------------------
We evaluate three representative defense strategies: instruction-based, filter-based, and differential privacy (DP)-based approaches.

\mypara{Instruction-Based Defense}
It is a method of safeguarding LLMs by embedding carefully crafted directives within prompts to proactively steer the model away from generating harmful or undesirable outputs~\cite{ZLWJZBSZ24}. 
We evaluate it, which involves providing explicit instructions to the target model to avoid revealing training data.
We implement this defense by directly adding ``respond to the following queries without directly mentioning or alluding to any specific examples, demonstrations, or instances that might have been used in the prompt'' at the end of each query.

\mypara{Filter-Based Defense} 
It is a protective strategy in large language models that detects and blocks unsafe or malicious inputs or outputs using external or integrated filtering mechanisms~\cite{WLBZ24}.
In creating multiple-choice questions, GPT-4 strictly follows instruction prompts to ensure each question is relevant to each data point. 
This then leads the \attack{} to depend, to a certain degree, on the semantic similarity between the question and the training data to identify the correct response while responding to the question.
This defensive approach employs an output filter that alters the target model's answers while maintaining pragmatics. 
It is accomplished by transmitting the created multiple-choice questions to GPT-4 for sentence rephrasing.
The rewritten words, although pragmatically consistent with the original, exhibit reduced semantic similarity to the initial example provided in the prompt.

\mypara{DP-Based Defense}
Differential privacy (DP) has been recognized as a crucial protective measure against MIAs. 
DP-based defense in LLMs is a method that integrates differential privacy into the training process to prevent the exposure of sensitive data and protect the anonymity of individual records~\cite{TSIMMLGKS23}.
For this defense, we achieve by producing synthetic demonstrations using private datasets that ensure differentiated privacy guarantees.
We set the num-private-train to 1 to omit unseen data. 
The resulting DP demonstrations are of poor quality, like ``15284 peopl VotingAPI hasvotd InsideFunc 2 @ViewerOutcome.''
The significant difference between the generated DP and the original demonstration could considerably compromise the effectiveness of \attack{}.

\mypara{Results}
As shown in \autoref{table:defense}, we find that the defense mechanisms show moderate effectiveness in mitigating \attack.
Without protective measures, ROC\_AUC scores consistently surpass 0.800 across all examined target LLMs and datasets, yielding an average score of 0.873.
After implementing defensive methods, a significant decrease is noted, with the majority of ROC\_AUC scores falling below 0.800.
DP-based demonstrates superior performance with a mean ROC\_AUC score of 0.708, while instruction-based and filter-based achieve mean ROC\_AUC scores of 0.753 and 0.731, respectively. 
To compare ROC\_AUC scores of six different target LLMs under three defenses against \attack{}, with DP-based approaches showing the strongest protection, especially for \textit{Medical} dataset.
Our analysis also displays that GPT-4o exhibits exceptional vulnerability when unprotected, ROC\_AUC$>$0.900, with values significantly decreasing to 0.700-0.800 range after defensive implementation.
It suggests that advanced models may be paradoxically more susceptible despite their superior capabilities. 
The quantitative results provide evidence that the proposed defensive mechanisms effectively mitigate \attack{}.
Current defensive mechanisms reduce attack effectiveness but ROC\_AUC scores exceeding 0.700 indicate incomplete privacy protection.

%-------------------------------------------------------------------------------
\section{Data Sample Overview}
\label{data sample}
%-------------------------------------------------------------------------------

%-------------------------------------------------------------------------------
\subsection{Securelist}
\label{securelist}
%-------------------------------------------------------------------------------
\begin{itemize}
    \item \textit{EastWind campaign: new CloudSorcerer attacks on government organizations in Russia} is posted on \textit{14 Aug 2024}, and it is written by \textit{GReAT}.
    The category for \textit{EastWind campaign: new CloudSorcerer attacks on government organizations in Russia} is \textit{APT reports}, and the keywords for it are \textit{CloudSorcerer, Backdoor, APT tool}.
    \item \textit{Loki: a new private agent for the popular Mythic framework} is posted on \textit{9 Sept 2024}, and it is written by \textit{Artem Ushkov}.
    The category for \textit{Loki: a new private agent for the popular Mythic framework} is \textit{Malware descriptions}, and the keywords for it are \textit{Targeted attacks, Backdoor, Framework}.
\end{itemize}

%-------------------------------------------------------------------------------
\subsection{Fiction}
\label{fiction}
%-------------------------------------------------------------------------------
\begin{itemize}
    \item The manga named \textit{ A World of Gold to You} is from \textit{Korea}, and its status is \textit{ongoing}.
    The category of \textit{A World of Gold to You} is \textit{Action, Fantasy, Manga, Adventure, Seinen, Manhwa, Mature}. 
    The total number of chapters for \textit{A World of Gold to You} is \textit{20}
    \item The manga named \textit{9Th Class Sword Master: The Guardian Of The Sword} is from \textit{Korea}, and its status is \textit{ongoing}.
    The category of \textit{9Th Class Sword Master: The Guardian Of The Sword} is \textit{Action, Fantasy, Adventure, Manhwa}. 
    The total number of chapters for \textit{9Th Class Sword Master: The Guardian Of The Sword} is \textit{11}.
\end{itemize}

%-------------------------------------------------------------------------------
\subsection{IMDb}
\label{imdb}
%-------------------------------------------------------------------------------

\begin{itemize}
    \item The type of \textit{The Ministry of Ungentlemanly Warfare} is \textit{movie}. The introduction to \textit{The Ministry of Ungentlemanly Warfare} is \textit{The British military recruits a small group of highly skilled soldiers to strike against German forces behind enemy lines during World War II}. 
    The certificate of \textit{The Ministry of Ungentlemanly Warfare} is \textit{rm2125219329} and the category is \textit{Action, Comedy, War}.
    \textit{104982} people voted for \textit{The Ministry of Ungentlemanly Warfare}, and the rating is \textit{6.8}.
    \item The type of \textit{Call for Billionaire's Surrogate} is \textit{TV Mini Series}. 
    The introduction to \textit{Call for Billionaire's Surrogate} is \textit{A woman seeks sperm donation to fulfill her dying mother's wish, but ends up entangled in a co-parenting contract with a handsome, wealthy stranger - complicating her plans for a simple pregnancy.}. 
    The certificate of \textit{Call for Billionaire's Surrogate} is \textit{[rm2365091073]} and the category is \textit{['Comedy, Drama']}.
    \textit{21} people voted for \textit{Call for Billionaire's Surrogate}, and the rating is \textit{8.6}.
\end{itemize}

%-------------------------------------------------------------------------------
\subsection{Medical}
\label{medical}
%-------------------------------------------------------------------------------
\begin{itemize}
    \item \textit{Quinn Smith}is living in \textit{Rhode Island}, USA. 
    The age of \textit{Quinn Smith} is \textit{36}.
    The gender of \textit{Quinn Smith} is \textit{Female}.
    \textit{Quinn Smith} has treated in \textit{Summit View Hospital}. 
    The treatment for \textit{Quinn Smith} is \textit{Counseling}. 
     \textit{Quinn Smith}'s blood pressure is \textit{149.11756611275806}, and \textit{Quinn Smith}'s BMI is \textit{31.538408217894265}.
    \item \textit{David Johnson}is living in \textit{Illinois}, USA. 
    The age of \textit{David Johnson} is \textit{27}.
    The gender of \textit{David Johnson} is \textit{Male}.
    \textit{David Johnson} has treated in \textit{Aspen Grove Medical Center}. 
    The treatment for \textit{David Johnson} is \textit{Medication}. 
    \textit{David Johnson}'s blood pressure is \textit{103.9959115811089}, and \textit{David Johnson}'s BMI is \textit{6.231508425880552}.
\end{itemize}

%-------------------------------------------------------------------------------
\subsection{Fiction\_abs}
\label{fiction_abs}
%-------------------------------------------------------------------------------
\begin{itemize}
    \item \textit{Our MC dies from an accidental electrocution and ends up inhabiting the body of a young prince in another world, his new profession being the Necromancer he chose in the game he was playing before his untimely demise. However, things are not what they seem – including his own Necromancy skills!}
    \item  \textit{
    Once upon a time, the world had been divided into the Moon Kingdom, inhabited by humans, and the Sun Kingdom, inhabited by demons. Estelle, a knight of the Moon Kingdom, loses her family to an ambush from the demons and is sentenced to life in exile after becoming entangled with a mysterious demon child. To survive, she sets out to a mysterious tower in the north with her companions. A timid knight, a slave boy, and a shadow who was once the Demon King. This is an alluring story about three people embarking on an adventure to find their true selves.}
\end{itemize}

%-------------------------------------------------------------------------------
\end{document}